\documentclass[prd,papera4]{revtex4}
\usepackage{amsmath,graphicx}

\begin{document}

\title{Upper and lower limits on the number of bound states in a central
potential}
\author{Fabian \surname{Brau}}
\email[E-mail: ]{fabian.brau@umh.ac.be}
\affiliation{Service de Physique G\'en\'erale et de Physique des Particules
El\'ementaires, Groupe de Physique Nucl\'eaire Th\'eorique, Universit\'e de
Mons-Hainaut, Mons, Belgique}
\author{Francesco \surname{Calogero}}
\email[E-mail: ]{francesco.calogero@roma1.infn.it}
\affiliation{Dipartimento di Fisica, Universit\`a di Roma ``La Sapienza" and Istituto
Nazionale di Fisica Nucleare, Sezione di Roma, Rome, Italy}
\date{\today}

\begin{abstract}
In a recent paper new upper and lower limits were given, in the context of
the Schr\"{o}dinger or Klein-Gordon equations, for the number $N_{0}$ of
S-wave bound states possessed by a monotonically nondecreasing central
potential vanishing at infinity. In this paper these results are extended to
the number $N_{\ell}$ of bound states for the $\ell$-th partial wave, and
results are also obtained for potentials that are not monotonic and even
somewhere positive. New results are also obtained for the case treated
previously, including the remarkably neat \textit{lower} limit $N_{\ell}\geq
\left\{\left\{\left[\sigma /(2\ell+1)+1\right]/2\right\}\right\}$ with $%
\sigma =\left(2/\pi\right) \underset{0\leq r<\infty }{\max}\left[r\left\vert
V(r)\right\vert ^{1/2}\right]$ (valid in the Schr\"{o}dinger case, for a
class of potentials that includes the monotonically nondecreasing ones),
entailing the following \textit{lower} limit for the total number $N$ of
bound states possessed by a monotonically nondecreasing central potential
vanishing at infinity: $N\geq
\left\{\left\{\left(\sigma+1\right)/2\right\}\right\}\left\{\left\{
\left(\sigma+3\right)/2\right\} \right\}/2$ (here the double braces denote
of course the integer part).
\end{abstract}

\maketitle

\section{INTRODUCTION AND MAIN RESULTS}

\label{sec1}

In a previous paper \cite{BC} new upper and lower limits were provided for
the number $N_{0}$ of S-wave bound states possessed, in the framework of the
Schr\"{o}dinger or Klein-Gordon equations, by a central potential $V(r)$
vanishing at infinity and having the property to yield a nowhere repulsive
force, so that, for all (nonnegative) values of the radius $r$, 
\begin{equation}
V^{\prime }(r)\geq 0,  \label{V1}
\end{equation}%
hence 
\begin{equation}
-V(r)=\left\vert V(r)\right\vert .  \label{V2}
\end{equation}%
In (\ref{V1}), and always below, appended primes signify of course
differentiation with respect to the radius $r$. The main purpose of the
present paper is to extend the results of \cite{BC} to higher partial waves,
namely to provide \textit{new} upper and lower limits for the number $%
N_{\ell }$ of $\ell $-wave bound states possessed by a central potential $%
V(r)$. Here and always below $\ell $ is of course the angular momentum
quantum number (a nonnegative integer). For simplicity we restrict attention
here to the Schr\"{o}dinger case, since the extension of the results to the
Klein-Gordon case is essentially trivial, see \cite{BC}. As in \cite{BC} we
assume the potential to be finite for $0<r<\infty $, to vanish at infinity
faster than the inverse square of $r$, 
\begin{equation}
\underset{r\rightarrow \infty }{\lim }\left[ r^{2+\varepsilon }V(r)\right]
=0,  \label{VatInf}
\end{equation}%
and, unless otherwise specified, not to diverge at the origin faster than
the inverse square of $r$, 
\begin{equation}
\underset{r\rightarrow 0}{\lim }\left[ r^{2-\varepsilon }V(r)\right] =0.
\label{VatZero}
\end{equation}%
Here of course $\varepsilon $ denotes some positive quantity, $\varepsilon >0
$. Moreover, in the following the \textquotedblleft
monotonicity\textquotedblright\ property (\ref{V1}) is generally replaced by
the less stringent condition 
\begin{subequations}
\label{CondV1}
\begin{equation}
-\frac{V^{\prime }(r)}{V(r)}+\frac{4\,\ell }{r}\geq 0,  \label{CondV1a}
\end{equation}%
or equivalently 
\begin{equation}
\left[ V(r)r^{-4\ell }\right] ^{\prime }\geq 0,  \label{CondV1b}
\end{equation}%
which is of course automatically satisfied by monotonic potentials vanishing
at infinity, see (\ref{V1}) and (\ref{V2}); and we also obtain results for
potentials that do not necessarily satisfy for all values of $r$ the
\textquotedblleft monotonicity\textquotedblright\ condition (\ref{CondV1})
and possibly not even the \textquotedblleft negativity\textquotedblright\
property (\ref{V2}). In any case the properties of the potential required
for the validity of the various results reported below will be specified in
each case.

In the process of deriving the results presented below we also uncovered
some \textit{new} neat limits (such as those reported in the Abstract) which
are as well applicable in the S-wave case and are different from those given
in \cite{BC}. These results therefore extend those presented in \cite{BC}.

From the upper and lower limits for $N_{\ell }$ one can obtain upper and
lower limits for the total number, 
\end{subequations}
\begin{equation}
N=\sum_{\ell =0}^{L}\left( 2\ell +1\right) \,N_{\ell },  \label{TotNum}
\end{equation}%
of bound states possessed by the potential $V(r)$; the upper limit, $L$, of
the sum in the right-hand side of this formula, (\ref{TotNum}), is of course
the largest value of $\ell $ for which the potential $V(r)$ possesses bound
states. It is well known that the conditions (\ref{VatInf}) and (\ref%
{VatZero}) are sufficient to guarantee that both $L$ and $N$ are finite. 
\textit{New} upper and lower limits on the values of the maximal angular
momentum quantum number $L$ for which bound states do exist are also
exhibited below, as well as \textit{new} upper and lower limits on the total
number of bound states $N$. Note that we are assuming, see (\ref{TotNum}),
to work in the (ordinary) three-dimensional world, with spherically
symmetrical potentials.

As in \cite{BC}, we begin below with a terse review of \textit{known}
results, and we then exhibit our \textit{new} upper and lower limits and
briefly outline their main features. A more detailed discussion of the
properties of these \textit{new} limits, including tests for various
potentials of their cogency (compared with that of previously known limits),
are then presented in Section \ref{sec2}. The proofs of our results are
given in Section \ref{sec3}, and some final remarks in Section \ref{sec4}.

\subsection{Units and preliminaries}

\label{sec1.1}

We use the standard quantum-mechanical units such that $2m=\hbar =1$, where $%
m$ is the mass of the particle bound by the central potential $V(r)$. This
entails that the potential $V(r)$ has the dimension of an inverse square
length, hence the following two quantities are dimensionless: 
\begin{equation}
S=\frac{2}{\pi }\int\nolimits_{0}^{\infty }dr\,\left[ -V^{(-)}(r)\right]
^{1/2},  \label{S}
\end{equation}
\begin{equation}
\sigma =\frac{2}{\pi }\,\underset{0\leq r<\infty }{\max }\left\{ r\,\left[
-V^{(-)}(r)\right] ^{1/2}\right\} .  \label{sigma}
\end{equation}
Here, and always below, $V^{(-)}(r)$ denotes the potential that obtains from 
$V(r)$ by setting to zero its positive part, 
\begin{equation}
V^{(-)}(r)=V(r)\,\theta \left[ -V(r)\right] .  \label{Vminus}
\end{equation}
Here, and always below, $\theta (x)$ is the standard step function, $%
\theta(x)=1$ if $x\geq 0$, $\theta (x)=0$ if $x<0$.

These quantities, $S$ and $\sigma$, play an important role in the following.
The motivation for inserting the $(2/\pi)$ prefactor in these definitions is
to make neater some of the formulas given below.

As for the $\ell $-wave radial Schr\"{o}dinger equation, in these standard
units it reads 
\begin{equation}
u_{\ell }^{\prime \prime }(\kappa ;r)=\left[ \kappa ^{2}+V(r)+\frac{\ell
(\ell +1)}{r^{2}}\right] \,u_{\ell }(\kappa ;r).  \label{SchEq}
\end{equation}%
The eigenvalue problem based on this ordinary differential equation (ODE)
characterizes the (moduli of the) $\ell $-wave bound-state energies, $\kappa
^{2}=\kappa _{\ell ,n}^{2}$, via the requirement that the corresponding
eigenfunctions, $u_{\ell }(\kappa _{\ell ,n};r)$ vanish at the origin, 
\begin{equation}
u_{\ell }(\kappa _{\ell ,n};0)=0,  \label{BounCondZero}
\end{equation}%
and be normalizable, hence vanish at infinity, 
\begin{equation}
\underset{r\rightarrow \infty }{\lim }\left[ u_{\ell }(\kappa _{\ell ,n};r)%
\right] =0.  \label{BounCondInf}
\end{equation}

It is well known that the conditions (\ref{VatInf}) and (\ref{VatZero}) on
the potential $V(r)$ are sufficient to guarantee that the (singular)
Sturm-Liouville problem characterized by the ODE (\ref{SchEq}) with the
boundary conditions (\ref{BounCondZero}) and (\ref{BounCondInf}) have a
finite (possibly vanishing) number of discrete eigenvalues $\kappa
_{\ell,n}^{2}$. To count them one notes that for sufficiently large (for
definiteness, positive) values of $\kappa$ the solution $u_{\ell }(\kappa;r)$
of the radial Schr\"{o}dinger equation (\ref{SchEq}) with the boundary
condition (\ref{BounCondZero}) (which characterizes the solution uniquely up
to a multiplicative constant) has no zeros in the interval $0<r<\infty$ and
diverges as $r\rightarrow \infty $ (proportionally to $\exp(\kappa r)$),
because for sufficiently large values of $\kappa$ the quantity in the square
bracket on the right-hand side of the radial Schr\"{o}dinger equation (\ref%
{SchEq}) is positive for all values of $r$, hence the solution $%
u_{\ell}(\kappa ;r)$ of this second-order ODE, (\ref{SchEq}), is everywhere
convex. Let us then imagine to decrease gradually the value of the positive
constant $\kappa$ so that the quantity in the square bracket in the
right-hand side of (\ref{SchEq}) becomes negative in some region(s) (for
this to happen the potential $V(r)$ must be itself negative in some
region(s), this being of course a necessary condition for the existence of
bound states), entailing that the solution $u_{\ell}(\kappa ;r)$ becomes
concave in that region(s). For some value, say $\kappa =\kappa _{\ell,1}$,
the solution $u_{\ell}(\kappa _{\ell,1};r)$ may then have a zero at $%
r=\infty $, namely vanish as $r\rightarrow \infty$ (proportionally to $%
\exp(-\kappa_{\ell ,1}\,r)$), thereby satisfying the boundary condition at
infinity (\ref{BounCondInf}) hence qualifying as a bound-state wave function
and thereby entailing that $\kappa _{\ell ,1}^{2}$ is the (modulus of the)
binding energy of the first (the most bound) $\ell $-wave state associated
with the potential $V(r)$. If one decreases $\kappa $ below $\kappa _{\ell
,1}$, the zero will then enter (from the right) the interval $0<r<\infty ,$
occurring, say, at $r=r_{\ell,1}(\kappa )$ (namely $u_{\ell }\left[ \kappa
;r_{\ell ,1}(\kappa )\right]=0$ with $0<r_{\ell ,1}(\kappa )<\infty $),
since the effect of decreasing $\kappa $, by decreasing the value of the
quantity in the square bracket in the right-hand side of the radial Schr\"{o}%
dinger equation (\ref{SchEq}), is to make the solution $u_{\ell }(\kappa ;r)$
more concave, hence to move its zeros to smaller values of $r$ (towards the
left on the positive real line $0<r<\infty $). Continuing the process of
decreasing $\kappa $, for $\kappa=\kappa _{\ell ,2}$ a second zero of the
solution $u_{\ell }(\kappa ;r)$ may appear at $r=\infty ,$ entailing that
this solution, $u_{\ell }(\kappa_{\ell ,2};r)$, satisfies again the boundary
condition at infinity (\ref{BounCondInf}), hence qualifies as a bound-state
wave function, implying that $\kappa _{\ell ,2}^{2}$ is the (modulus of the)
binding energy of the next most bound $\ell $-wave state associated with the
potential $V(r)$. The process can then be continued, yielding a sequence of
decreasing (in modulus) binding energies $\kappa _{\ell ,n}^{2}$ with $%
n=1,2,\ldots,N_{\ell }$. Correspondingly, the solution $u_{\ell }(\kappa ;r)$
of the radial Schr\"{o}dinger equation (\ref{SchEq}) characterized by the
boundary condition (\ref{BounCondZero}) shall have, for $\kappa _{\ell
,n-1}>\kappa>\kappa _{\ell ,n}$, $n-1$ zeros in the interval $0<r<\infty $.
The process of decreasing the parameter $\kappa $ we just described ends
when this parameter reaches the value $\kappa =0$, and it clearly entails
that the number of zeros $r_{\ell ,n}(0)$ of the zero-energy solution $%
u_{\ell }(0;r)$ of the radial Schr\"{o}dinger equation (\ref{SchEq})
characterized by the boundary condition (\ref{BounCondZero}) coincides with
the number $N_{\ell }$ of bound states possessed by the potential $V(r)$
(namely: $u_{\ell }\left[0;r_{\ell ,n}(0)\right] =0$ with $0<r_{\ell
,1}<r_{\ell ,2}<\ldots <r_{\ell,N_{\ell }-1}<r_{\ell ,N_{\ell }}<\infty $).

Hence in the following -- as indeed in \cite{BC} -- in order to obtain upper
and lower limits on the number $N_{\ell }$ of $\ell $-wave bound states we
focus on obtaining upper and lower limits on the number $N_{\ell }$ of zeros
of the zero-energy solution $u_{\ell }(0;r)$ of the radial Schr\"{o}dinger
equation (\ref{SchEq}) characterized by the boundary condition (\ref%
{BounCondZero}) (for notational simplicity, these zeros will be hereafter
denoted as $z_{n}$, and the zero-energy solution of the radial Schr\"{o}%
dinger equation (\ref{SchEq}) characterized by the boundary condition (\ref%
{BounCondZero}) as $u(r),$ namely $u(r)\equiv u_{\ell }(0;r)$ with $%
u(z_{n})=0,$ $0<z_{1}<\ldots <z_{N_{\ell }}<\infty $: see Section \ref{sec3}%
). Let us moreover emphasize that here, and throughout this paper, we ignore
the marginal possibility that the potential $V(r)$ under consideration
possess a \textquotedblleft zero-energy\textquotedblright bound state,
namely that $u(r)$ vanish as $r\rightarrow \infty $ (for $\ell >0)$, or tend
to a constant value in the S-wave case; namely, we assume $z_{N_{\ell
}}<\infty $, because keeping this possibility into account would force us to
go several times into cumbersome details, the effort to do so being clearly
out of proportion with the additional clarification gained.

In the following subsections we briefly review the \textit{known}
expressions, in terms of a given central potential $V(r)$, of upper and
lower limits on the number $N_{\ell }$ of $\ell $-wave bound states, and
also on the maximum value, $L$, of the angular momentum quantum number for
which bound states do exist, as well as on the total number $N$ of bound
states, see (\ref{TotNum}); and we also present our \textit{new} upper and
lower limits on these quantities.

Before listing these upper and lower limits let us note that an immediate
hunch on the accuracy of these limits for strong potentials may be obtained
via the introduction of a (dimensionless, positive) \textquotedblleft
coupling constant\textquotedblright\ $g$ by setting 
\begin{equation}
V(r)=g^{2}\,v(r),  \label{gDef}
\end{equation}
where $v(r)$ is assumed to be independent of $g,$ and by recalling that, at
large $g$, $N_{\ell }$ grows proportionally to $g$ \cite{CalBook}, 
\begin{subequations}
\label{NsubLasy}
\begin{equation}
N_{\ell }\sim g\quad \text{as\quad }g\rightarrow \infty ,  \label{NsubLasy1}
\end{equation}
indeed \cite{Cha} 
\begin{equation}
N_{\ell }\approx \frac{1}{\pi }\int_{0}^{\infty }dr\,\left[-V^{(-)}(r)\right] ^{1/2}\quad \text{as\quad }g\rightarrow \infty .
\label{NsubLasy2}
\end{equation}
Here, and always below, we denote with the symbols $\approx $ respectively $%
\sim $ asymptotic equality respectively proportionality.

The analogous asymptotic behaviors of $L$ and of $N$ read 
\end{subequations}
\begin{subequations}
\label{Lasy}
\begin{equation}
L\sim g,\quad \text{as\quad }g\rightarrow \infty ,  \label{Lasy1}
\end{equation}
indeed \cite{BS} 
\begin{equation}
L\approx \underset{0\leq r<\infty }{\max }\left\{ r\left[-V^{(-)}(r)\right] ^{1/2}\right\}\quad \text{as\quad }g\rightarrow \infty ,
\label{Lasy2}
\end{equation}
and 
\end{subequations}
\begin{subequations}
\label{NpropG3}
\begin{equation}
N\sim g^{3}\quad \text{as\quad }g\rightarrow \infty  \label{NpropG3a}
\end{equation}
indeed \cite{AM1} 
\begin{equation}
N\approx \frac{2}{3\pi }\int_{0}^{\infty }dr\,r^{2}\,\left[-V^{(-)}(r)\right]^{3/2}\quad \text{as\quad }g\rightarrow \infty.
\label{NpropG3b}
\end{equation}

\subsection{Limits defined in terms of global properties of the potential
(i. e., involving integrals over the potential)}
\label{sec1.2}

In this subsection we only consider results which can be formulated in terms
of integrals over the potential $V(r)$, possibly raised to a power, see
below. We firstly review tersely \textit{known }upper and lower limits on
the number $N_{\ell }$ of bound states, as well as \textit{known} upper and
lower limits on the maximum value $L$ for which bound states do exist; and
we then provide \textit{new} upper and lower limits for both $N_{\ell }$ and 
$L$.

The earliest upper limit of this kind on the number $N_{\ell }$ of $\ell $%
-wave bound states is due to V. Bargmann \cite{barg} (and then also
discussed by J. Schwinger \cite{sch}), and we hereafter refer to it as the BS%
$\ell$ upper limit:

\end{subequations}
\begin{equation}
\text{BS}\ell \text{:\quad }N_{\ell }<\frac{1}{2\ell +1}\int\nolimits_{0}^{
\infty }dr\,r\,\left[- V^{(-)}(r)\right] .  \label{BS1}
\end{equation}

\textit{Remark}. In writing this upper limit we have used the strict
inequality sign; \textit{we will follow this rule in all the analogous
formulas we write hereafter}. Let us repeat that in this manner we
systematically ignore the possibility that a potential possess exactly the
number of bound states given by the (upper or lower) limit expression being
displayed (which in such a case would have to yield an integer value), since
this would correspond to the occurrence of a \textquotedblleft zero-energy
bound state\textquotedblright\ (in the S-wave case) or a \textquotedblleft
zero-energy resonance\textquotedblright\ (in the higher-wave case) -- a
marginal possibility we believe can be ignored without significant loss of
generality.

Since the right-hand side of this inequality, (\ref{BS1}), grows
proportionally to $g^{2}$ (see (\ref{gDef})) rather than $g$ (see (\ref%
{NsubLasy})) as $g$ diverges, for strong potentials possessing many bound
states this upper limit, (\ref{BS1}), is generally very far from the exact
value. It clearly implies the following upper limit on $L$: 
\begin{equation}
\text{BSL:\quad }L<L_{\text{BSL}}^{(+)}=-\frac{1}{2}+\frac{1}{2}%
\int\nolimits_{0}^{\infty }dr\,r\, \left[- V^{(-)}(r)\right] .
\label{BS2}
\end{equation}
The right-hand side of this inequality also grows proportionally to $g^{2}$
(see (\ref{gDef})) rather than to $g$ (see (\ref{Lasy})) as $g$ diverges,
hence for strong potentials possessing many bound states this upper limit, (%
\ref{BS2}), is also generally very far from the exact value. The limit BS$%
\ell$ is however best possible, namely there is a potential $V(r)$, 
\begin{equation}
V(r)=g\,R^{-1}\sum_{n=1}^{N_{\ell }}\alpha _{n}\,\delta (r-\beta _{n}\,R)
\end{equation}
(with an appropriate assignment of the $2N_{\ell}$ dimensionless constants $%
\alpha _{n}$ and $\beta _{n},$ depending of course on $g$, on $\ell$ and on $%
N_{\ell }$) that possesses $N_{\ell}$ $\ell$-wave bound states and for which
the right-hand side of (\ref{BS1}) takes a value arbitrarily close to $%
N_{\ell}$.

The next upper limit we report is due to K. Chadan, A. Martin and J. Stubbe 
\cite{CMS}, and we denote it as CMS. It holds only for potentials that
satisfy the monotonicity condition (\ref{V1}), and it reads (see (\ref{S})): 
\begin{equation}
\text{CMS:\quad }N_{\ell }<S+1-\sqrt{1+\left(\frac{2}{\pi }\right)^{2}\,\ell (\ell +1)}.
\label{CMS}
\end{equation}%
A less stringent but neater version \cite{CMS} of this upper limit, which we
denote as CMSn, reads 
\begin{equation}
\text{CMSn:\quad }N_{\ell }<S+1-\frac{2\ell +1}{\pi }.  \label{CMSn}
\end{equation}%
Clearly this inequality, (\ref{CMSn}), entails the following neat upper
limit on $L$, which we denote as CMSL: 
\begin{equation}
\text{CMSL:}\quad L<L_{\text{CMSL}}^{(+)}=\frac{\pi }{2}S-\frac{1}{2}.
\label{CMSL}
\end{equation}%
A more stringent but less neat upper limit on $L$, which we do not write,
can of course be obtained from the CMS upper limit (\ref{CMS}).

The next upper limit we report is immediately implied by a result due to A.
Martin \cite{AM}, and we denote it as M$\ell $. It reads 
\begin{equation}
\text{M}\ell \text{:\quad }N_{\ell }<\left[ \int\nolimits_{0}^{\infty
}dr\,r^{2}\,V_{\ell ,\text{eff}}^{(-)}(r)\int\nolimits_{0}^{\infty
}dr\,V_{\ell ,\text{eff}}^{(-)}(r)\right] ^{1/4},  \label{M}
\end{equation}%
with $V_{\ell ,\text{eff}}^{(-)}(r)$ being the negative part of the
\textquotedblleft effective $\ell $-wave potential\textquotedblright\ (see (%
\ref{SchEq})) 
\begin{equation}
V_{\ell ,\text{eff}}(r)=V(r)+\frac{\ell (\ell +1)}{r^{2}}.  \label{Veff}
\end{equation}

Finally, the last two upper limits of this type we report are due to V.
Glaser, H. Grosse, A. Martin and W. Thirring \cite{GGMT}, and again to K.
Chadan, A. Martin and J. Stubbe \cite{CMS2}, and we denote them as GGMT and
CMS2. The first of these upper limits reads 
\begin{subequations}
\begin{equation}
\text{GGMT:\quad }N_{\ell
}<(2\ell+1)^{1-2p}\,C_{p}\,\int\nolimits_{0}^{\infty }\frac{dr}{r}\,\left[%
-r^{2}\, V^{(-)}(r)\right] ^{p}  \label{GGMT1}
\end{equation}
with

\begin{equation}
C_{p}=\frac{(p-1)^{p-1}\,\Gamma (2p)}{p^{p}\,\Gamma ^{2}(p)},  \label{GGMT2}
\end{equation}
and the restriction $p\geq 1$. This upper limit GGMT is however always
characterized by an unsatisfactory dependence on $g$ as $g\rightarrow \infty$
(see (\ref{gDef})): the right-hand side of (\ref{GGMT1}) is proportional to $%
g^{2p}$ with $p\geq 1$ rather than to $g$ (see (\ref{gDef}) and (\ref%
{NsubLasy})), hence it always yields a result far from the exact value for
strong potentials possessing many bound states.

The second of these upper limits reads

\label{CMS2} 
\end{subequations}
\begin{subequations}
\begin{equation}
\text{CMS2:\quad }N_{\ell }<(2\ell +1)^{1-2p}\,\tilde{C}_{p}\,\int%
\nolimits_{0}^{\infty }\frac{dr}{r}\left\vert r^{2}\,V(r)\right\vert ^{p},
\label{CMS2a}
\end{equation}%
with 
\begin{equation}
\tilde{C}_{p}=p\,(1-p)^{p-1},  \label{CMS2b}
\end{equation}%
with the restriction $1/2\leq p<1,$ and it is valid provided the potential
is nowhere positive, see (\ref{V2}), and moreover satisfies for all values
of $r,$ $0\leq r<\infty ,$ the relation

\end{subequations}
\begin{equation}
\left\{ r^{1-2p}\left[ -V(r)\right] ^{1-p}\right\} ^{\prime }\leq 0.
\label{Condp}
\end{equation}%
Note that the right-hand side of (\ref{CMS2a}) features the correct power
growth proportional to $g,$ see (\ref{gDef}) and (\ref{NsubLasy}), only if $%
p=1/2$, in which case the condition (\ref{Condp}) is equivalent to the
condition (\ref{V1}) but the upper limit CMS2 then reads (see (\ref{CMS2})
and (\ref{S})) 
\begin{equation}
N_{\ell }<2^{-3/2}\pi \,S,
\end{equation}%
hence it is analogous, but less stringent (since $2^{-3/2}\pi \cong 1.11>1$%
), than the limit CC, 
\begin{equation}
\text{CC:\quad }N_{\ell }<S,  \label{CC}
\end{equation}%
(see eq. (1.4) of \cite{BC}; of course this limit, valid for S-waves, is 
\textit{a fortiori} valid for all partial waves -- albeit clearly not very
good for large $\ell $, and moreover off by a factor 2 in the asymptotic
limit of strong potentials, see (\ref{gDef}) and (\ref{NsubLasy2}) -- indeed
the main motivation for, and achievement of, the research reported in \cite%
{BC} was just to provide upper and lower limits to $N_{0}$ that do not have
this last defect).

Let us turn now to \textit{known} lower limits, always restricting our
consideration here to results which can be formulated in terms of integrals
over the potential $V(r)$, possibly raised to a power, see below.

Only one result of this kind seems to be previously known \cite{Cal1,CalBook}%
, and we will denote it as C$\ell$. It states (without requiring any
additional conditions on the potential other than (\ref{VatInf}) and (\ref%
{VatZero}) -- and even these conditions are sufficient but non necessary)
that

\begin{equation}
\text{C}\ell \text{:\quad }N_{\ell }>-\frac{1}{2}+\frac{1}{\pi }%
\,\max_{0\leq a<\infty }\left\{ \int_{0}^{\infty }dr\,\min \left[ a^{-1}\left(\frac{r}{a}\right)^{2\ell },-a\,V(r)\left(\frac{r}{a}\right)^{-2\ell }\right] \right\} .
\label{loCal}
\end{equation}%
In this formula, and hereafter, the notation $\min \left[ x,y\right] $
signifies $x$ if $x\leq y$, $y$ if $y\leq x.$ Let us now assume that the
equation 
\begin{equation}
a^{-1}\left(\frac{r}{a}\right)^{2\ell }=-a\,V(r)\left(\frac{r}{a}\right)^{-2\ell }  \label{R(a)}
\end{equation}%
admits one and only one solution, say $r=R(a)$ (and note that validity for
all values of $r$ of the \textquotedblleft monotonicity
condition\textquotedblright\ (\ref{CondV1}) is sufficient to guarantee that
this is indeed the case), so that the lower limit (\ref{loCal}) can be
rewritten as follows: 
\begin{equation}
N_{\ell }>-\frac{1}{2}+\frac{1}{\pi }\,\underset{0\leq a<\infty }{\max }%
\left\{ \int_{0}^{R(a)}dr\,a^{-1}\left(\frac{r}{a}\right)^{2\ell }-\int_{R(a)}^{\infty
}dr\,a\,V(r)\left(\frac{r}{a}\right)^{-2\ell }\right\} ,  \label{ineq1}
\end{equation}%
where of course $r=R(a)$ is the solution of (\ref{R(a)}). It is then easy,
using (\ref{R(a)}), to calculate the maximum in the right-hand side of this
inequality (note that the first integral inside the braces in the right-hand
side of the above inequality, (\ref{ineq1}), is elementary) and to obtain
thereby the following lower limit, that we denote here as C$\ell $n:

\begin{subequations}
\label{Cl}
\begin{equation}
\text{C}\ell \text{n}:\quad N_{\ell }>-\frac{1}{2}+\frac{2}{\pi }\frac{\rho
\,\left\vert V(\rho )\right\vert ^{1/2}}{2\ell +1}  \label{Cla}
\end{equation}%
with the radius $\rho $ defined to be the solution of the following
equation: 
\begin{equation}
\rho \,V(\rho )=(2\ell +1)\int\nolimits_{\rho }^{\infty }dr\,\left( \frac{%
\rho }{r}\right) ^{2\ell }\,V(r).  \label{Clb}
\end{equation}%
This lower limit C$\ell $n presents the correct dependence on $g$, see (\ref%
{gDef}) and (\ref{NsubLasy}), since clearly $\rho $ does not depend on $g$.
It is best possible, and the potential that saturates it has the form \cite%
{Cal1, CalBook} 
\end{subequations}
\begin{subequations}
\label{VeqR4}
\begin{equation}
V(r)=-g^{2}R^{-2}\,\left(\frac{r}{R}\right)^{4\ell }\,\quad \text{for\quad }0\leq
r<\alpha \,R,  \label{VeqR4a}
\end{equation}%
\begin{equation}
V(r)=0\quad \text{for\quad }r\geq \alpha \text{\thinspace }R,  \label{VeqR4b}
\end{equation}%
with $R$ an arbitrary (of course \textit{positive}) radius and $\alpha $ a
dimensionless constant given by 
\end{subequations}
\begin{equation}
\alpha =\left[ \frac{\pi \,(2\ell +1)\,(N_{\ell }+\delta )}{g}\right]
^{1/(2\ell +1)},\quad 0\leq \delta <1/2.  \label{Alpha1}
\end{equation}

Let us now present our \textit{new} upper and lower limits on the number of
bound states possessed by the central potential $V(r)$. All these results
are proven in Section \ref{sec3}.

We begin with a \textit{new} upper limit on the number $N_{0}$ of S-wave
bound states, possessed by a central potential $V(r)$ that features the
following properties: it has two zeros, $V(r_{\pm })=0$ (with $r_{-}<r_{+}),$
it is positive for $r$ smaller than $r_{-}$, negative for $r$ in the
interval from $r_{-}$ to $r_{+}$, again positive for $r$ larger than $r_{+}$,

\begin{subequations}
\label{Vr+-}
\begin{equation}
V(r)>0\quad \text{for\quad }0\leq r<r_{-},  \label{Vr+-a}
\end{equation}
\begin{equation}
V(r)<0\quad \text{for\quad }r_{-}<r<r_{+},  \label{Vr+-b}
\end{equation}
\begin{equation}
V(r)>0\quad \text{for\quad }r_{+}<r<\infty .  \label{Vr+-c}
\end{equation}
Note that we do not exclude the possibility that the potential diverge (but
then to \textit{positive} infinity) at the origin; so, for the validity of
the result we now report, the condition (\ref{VatZero}) need not hold, and
indeed even the condition (\ref{VatInf}) can be forsaken, provided the
potential does vanish at infinity, $V(\infty )=0$. Indeed one option we
shall exploit below is to replace the potential $V(r)$ with $V_{\ell ,\text{%
eff}}(r)$, see (\ref{Veff}), and to thereby include in the present framework
the treatment of the $\ell $-wave case. On the other hand the assumption
that the potential have only two zeros and no more is made here for
simplicity; the extension to potentials having more than two zeros is
straightforward, but the corresponding results lack the neatness that
justifies their explicit presentation here (we trust any potential user of
our results who needs to apply them to the more general case of a potential
with more than two zeros will be able to obtain easily the relevant formulas
by extending the treatment of Section \ref{sec3}). Let us also note that the
following results remain valid (but may become trivial) if $r_{-}=0$ or $%
r_{+}=\infty $.

We denote as NUL1 (\textquotedblleft New Upper Limit no.
1\textquotedblright) this result: 
\end{subequations}
\begin{equation}
\text{NUL1:\quad }N_{0}<1+\frac{2}{\pi }\left\{
(r_{+}-r_{-})\int_{0}^{\infty}dr\,\left[ -V^{(-)}(r)\right] \right\} ^{1/2}.
\label{NUL1}
\end{equation}
It can actually be shown (see subsection \ref{sec3.7}) that this upper limit
NUL1 is generally less cogent than the upper limit NUL2, see (\ref{NUL2bis})
below; but it has the advantage over NUL2 of being simpler, and for this
reason it is nevertheless worthwhile to report it separately here.

Let us now report a \textit{new }lower limit on the number $N_{0}$ of S-wave
bound states that holds for potentials that satisfy the same conditions (\ref%
{Vr+-}), and that we denote as NLL1 (\textquotedblleft New Lower Limit no.
1\textquotedblright). It is actually a variation of the lower limit C$\ell $%
, see (\ref{loCal}), and it reads:

\begin{equation}
\text{NLL1:\quad }N_{0}>-1+\frac{1}{\pi }\, \underset{0\leq a<\infty }{\max }%
\left\{ \int_{0}^{\infty }dr\,\min \left( a^{-1},a\left[ -V^{(-)}(r)\right]%
\right) \right\} .  \label{NLL1}
\end{equation}

Of course, a less cogent but perhaps simpler version of this lower bound
reads

\begin{equation}
\text{NLL1n:}\quad N_{0}>-1+\frac{1}{\pi }\left\{ \int_{0}^{\infty}dr\, %
\left[ -V^{(-)}(r)\right] \right\} \,\left\{ \max \left[ -V^{(-)}(r)\right]%
\right\} ^{-1/2}  \label{NLL1n}
\end{equation}
(it clearly obtains from NLL1 by setting $a=\left\{ \max \left[ -V^{(-)}(r)%
\right] \right\} ^{-1/2}$).

These \textit{new }upper and lower limits become relevant to the number $%
N_{\ell }$ of $\ell $-wave bound states possessed by the central potential $%
V(r)$ via the replacement in the above inequalities, (\ref{NUL1}) and (\ref%
{NLL1n}), of $V(r)$ with $V_{\ell ,\text{eff}}(r)$, see (\ref{Veff}). Note
that, for a large class of central potentials $V(r)$ satisfying the
conditions (\ref{VatInf}) and (\ref{VatZero}), this effective $\ell $-wave
potential $V_{\ell ,\text{eff}}(r),$ especially for $\ell >0$, is indeed
likely to satisfy the conditions (see (\ref{Vr+-})) 
\begin{subequations}
\begin{equation}
V_{\ell ,\text{eff}}(r)=V(r)+\frac{\ell \,(\ell +1)}{r^{2}}>0\quad \text{%
for\quad }0\leq r<r_{-}^{(\ell )},
\end{equation}%
\begin{equation}
V_{\ell ,\text{eff}}(r)=V(r)+\frac{\ell \,(\ell +1)}{r^{2}}<0\quad \text{%
for\quad }r_{-}^{(\ell )}<r<r_{+}^{(\ell )},
\end{equation}%
\begin{equation}
V_{\ell ,\text{eff}}(r)=V(r)+\frac{\ell \,(\ell +1)}{r^{2}}>0\quad \text{%
for\quad }r_{+}^{(\ell )}<r<\infty ,
\end{equation}%
required for the validity of the upper and lower limits NUL1 and NLL1, see (%
\ref{NUL1}) and (\ref{NLL1}). We denote the \textit{new} upper and lower
limits obtained in this manner as NUL1$\ell $ and NLL1n$\ell $: 
\end{subequations}
\begin{equation}
\text{NUL1}\ell \text{:\quad }N_{\ell }<1+\frac{2}{\pi }\left\{
(r_{+}^{(\ell )}-r_{-}^{(\ell )})\left[ -\ell (\ell +1)\,\left( \frac{1}{%
r_{-}^{(\ell )}}-\frac{1}{r_{+}^{(\ell )}}\right) +\int_{r_{-}^{(\ell
)}}^{r_{+}^{(\ell )}}dr\,\left\vert V(r)\right\vert \right] \right\} ^{1/2},
\label{NUL1L}
\end{equation}%
\begin{eqnarray}
\text{NLL1n}\ell  &\text{:}&\quad N_{\ell }>-1+\frac{1}{\pi }\left[ -\ell
(\ell +1)\,\left( \frac{1}{r_{-}^{(\ell )}}-\frac{1}{r_{+}^{(\ell )}}\right)
+\int_{r_{-}^{(\ell )}}^{r_{+}^{(\ell )}}dr\,\left\vert V(r)\right\vert %
\right] \cdot   \notag \\
&&\cdot \left[ \underset{r_{-}^{(\ell )}<r<r_{+}^{(\ell )}}{\max }\left\vert
V(r)+\frac{\ell (\ell +1)}{r^{2}}\right\vert \right] ^{-1/2}.  \label{NLL1l}
\end{eqnarray}

Next, we report \textit{new }upper and lower limits on the number $N_{0}$ of
S-wave bound states somewhat analogous to those given in \cite{BC}, but
applicable to nonmonotonic potentials. As above, we restrict for simplicity
our consideration to potentials that satisfy the conditions (\ref{Vr+-}). We
do moreover, again for simplicity, require the potential $V(r)$ to possess
only one minimum, at $r=r_{\text{min}}$: 
\begin{subequations}
\label{V+-}
\begin{equation}
V(r)>0\quad \text{for\quad }0\leq r<r_{-},  \label{V+-a}
\end{equation}
\begin{equation}
V(r)<0,\,V^{\prime }(r)\leq 0\quad \text{for\quad }r_{-}<r\leq r_{\text{min}
},  \label{V+-b}
\end{equation}
\begin{equation}
V(r)<0,\,V^{\prime }(r)\geq 0\quad \text{for\quad }r_{\text{min}}\leq
r<r_{+},  \label{V+-c}
\end{equation}
\begin{equation}
V(r)>0\quad \text{for\quad }r_{+}<r<\infty .  \label{V+-d}
\end{equation}
We denote these \textit{new }upper, respectively lower, limits on the number 
$N_{0}$ of S-wave bound states as NUL2, respectively NLL2: 
\end{subequations}
\begin{equation}
\text{NUL2:\quad }N_{0}<1+\frac{S}{2}+\frac{1}{2\pi }\,\log \left[ \frac{%
-V^{(-)}(r_{\text{min}})}{M}\right] ,  \label{NUL2bis}
\end{equation}
\begin{equation}
\text{NLL2:\quad }N_{0}>-\frac{3}{2}+\frac{S}{2}-\frac{1}{2\pi }\,\log \left[
\frac{-V^{(-)}(r_{\text{min}})}{M}\right] ,  \label{NLL2bis}
\end{equation}
where $S$ is of course defined by (\ref{S}) and 
\begin{equation}  \label{Mdef}
M=\min \left[ -V^{(-)}(p),-V^{(-)}(q)\right]
\end{equation}
with the two radii $p$ and $q$ defined as the solutions of the following
equations: 
\begin{equation}
\int_{0}^{p}dr\,\left[ -V^{(-)}(r)\right] ^{1/2}=\frac{\pi }{2},  \label{P}
\end{equation}
\begin{equation}
\int_{q}^{\infty }dr\,\left[ -V^{(-)}(r)\right] ^{1/2}=\frac{\pi }{2},
\label{Q}
\end{equation}
and with the additional condition (which might rule out the applicability of
these limits to potentials possessing very few bound states, but which is
certainly satisfied by potentials that are sufficiently strong to possess
several bound states) 
\begin{equation}
p\leq r_{\text{min}}\leq q.  \label{PrminQ}
\end{equation}

As already mentioned above and explained in subsection \ref{sec3.7}, the
upper limit NUL2, see (\ref{NUL2bis}), is generally more cogent than the
upper limit NUL1, see (\ref{NUL1}), but it requires the additional
computation of the two radii $p$ and $q$.

Again, as above, \textit{new} limits (hereafter denoted NUL2$\ell $
respectively NLL2$\ell $) on the number $N_{\ell }$ of $\ell $-wave bound
states possessed by the central potential $V(r)$ are entailed by these
results via the replacement of $V(r)$ with $V_{\ell ,\text{eff}}(r),$ see (%
\ref{Veff}), so that the relevant formulas read as follows: 
\begin{subequations}
\label{V+-l}
\begin{equation}
V_{\ell ,\text{eff}}(r)>0\quad \text{for\quad }0\leq r<r_{-}^{(\ell )},
\label{V+-la}
\end{equation}
\begin{equation}
V_{\ell ,\text{eff}}(r)<0,\,V_{\ell ,\text{eff}}^{\prime }(r)\leq 0\quad 
\text{for\quad }r_{-}^{(\ell )}<r\leq r_{\text{min}}^{(\ell )},
\label{V+-lb}
\end{equation}
\begin{equation}
V_{\ell ,\text{eff}}(r)<0,\,V_{\ell ,\text{eff}}^{\prime }(r)\geq 0\quad 
\text{for\quad }r_{\text{min}}^{(\ell )}\leq r<r_{+}^{(\ell )},
\label{V+-lc}
\end{equation}
\begin{equation}
V_{\ell ,\text{eff}}(r)>0\quad \text{for\quad }r_{+}^{(\ell )}<r<\infty ;
\label{V+-ld}
\end{equation}
\end{subequations}
\begin{equation}
\text{NUL2}\ell \text{:\quad }N_{\ell }<1+\frac{1}{\pi }\,\int_{r_{-}^{(%
\ell)}}^{r_{+}^{(\ell )}}dr\,\left[ -V_{\ell ,\text{eff}}^{(-)}(r)\right]
^{1/2}+ \frac{1}{2\pi }\,\log \left\{ \frac{\left[ -V_{\ell ,\text{eff}%
}^{(-)}(r_{ \text{min}}^{(\ell )})\right] }{M}\right\},
\end{equation}
\begin{equation}
\text{NLL2}\ell \text{:\quad }N_{\ell }>-\frac{3}{2}+\frac{1}{\pi }%
\,\int_{r_{-}^{(\ell)}}^{r_{+}^{(\ell )}}dr\,\left[ -V_{\ell ,\text{eff}%
}^{(-)}(r)\right] ^{1/2}-\frac{1}{2\pi }\,\log \left\{ \frac{\left[ -V_{\ell
,\text{eff}}^{(-)}(r_{ \text{min}}^{(\ell )})\right] }{M}\right\};
\end{equation}
\begin{equation}
\int_{r_{-}^{(\ell )}}^{p^{(\ell )}}dr\,\left[ -V_{\ell ,\text{eff}}^{(-)}(r)%
\right] ^{1/2}=\frac{\pi }{2},
\end{equation}
\begin{equation}
\int_{q^{(\ell )}}^{r_{+}^{(\ell )}}dr\left[ -V_{\ell ,\text{eff}}^{(-)}(r)%
\right] ^{1/2}=\frac{\pi }{2};
\end{equation}
\begin{equation}
p^{(\ell )}\leq r_{\text{min}}^{(\ell )}\leq q^{(\ell )},
\end{equation}

\begin{equation}
M=\min \left[ -V_{\ell ,\text{eff}}^{(-)}(p^{(\ell )}),\,-V_{\ell ,\text{eff}%
}^{(-)}(q^{(\ell )})\right] .
\end{equation}

Let us now report another \textit{new} lower limit on the number $N_{\ell }$
of $\ell $-wave bound states applicable to nonmonotonic potentials. As
above, we restrict for simplicity our consideration to potentials that
satisfy the conditions (\ref{Vr+-}). We do moreover, again for simplicity,
require the potential $V(r)$ to possess only one minimum, at $r=r_{\text{min}%
}$, see (\ref{V+-}). We denote it by the acronym NLL3s: 
\begin{equation}
\text{NLL3s:\quad }N_{\ell }>-1+\frac{1}{\pi }\int_{0}^{s}dr\,\left[
-V^{(-)}(r)\right] ^{1/2}-\frac{1}{4\pi }\log \left\{ \frac{\left[
V^{(-)}(r_{\text{min}})\right] ^{2}}{V^{(-)}(p)V^{(-)}(s)}\right\} -\frac{%
\ell }{\pi }\log \left( \frac{s}{p}\right) ,  \label{NLL2s}
\end{equation}%
where the radius $p$ is defined by (\ref{P}) and $s$ is an arbitrary radius
(of course larger than $p$, $s>p$). The value of $s$ that yields the most
stringent limit is a (or the) solution of the equation 
\begin{equation}
s\,V^{\prime }(s)=4\,s\,\left\vert V(s)\right\vert ^{3/2}+4\ell \,V(s)
\end{equation}%
(since clearly for this value of $s$ the potential $V(r)$ is negative, $%
V(s)=-|V(s)|$, in this formula $V(s)$ could be replaced by $V^{(-)}(s)$ with
the condition $r_{-}<s<r_{+}$, see (\ref{Vr+-})).

A neater, if marginally less stringent, version of this lower limit NLL3s,
which we denote as NLL3, reads as follows: 
\begin{subequations}
\label{NLL2}
\begin{equation}
\text{NLL3:\quad }N_{\ell }>\nu -\frac{\ell }{\pi }\log \left( \frac{q}{p}%
\right)   \label{NLL2a}
\end{equation}%
where 
\begin{equation}
\nu =-\frac{3}{2}+\frac{1}{2}S-\frac{1}{4\pi }\log \left\{ \frac{\left[
V^{(-)}(r_{\text{min}})\right] ^{2}}{V^{(-)}(p)V^{(-)}(q)}\right\} .
\label{nu}
\end{equation}%
Here of course $p$ is defined as above, see (\ref{P}), while $q$ is defined
by the formula (\ref{Q}), and of course $S$ is defined by (\ref{S}). Here
and below, see (\ref{NLL2}) and (\ref{NLL2L}), as well as in all subsequent
formulas involving both $p$ and $q,$ see (\ref{P}) and (\ref{Q}), we always
assume validity of the inequality $q\geq p$, as is indeed generally the case
for any potential possessing enough bound states. [These results are also
valid if the potential has only one zero or no zero at all, and even if the
derivative of the potential never vanishes; in this latter case $r_{\text{min%
}}$ in (\ref{NLL2s}) and (\ref{nu}) must be replaced by $p$].

Clearly this lower limit, NLL3, see (\ref{NLL2}), implies the following 
\textit{new} lower limit $L_{\text{NLL3L}}^{(-)}$ on the largest value $L$
of the angular momentum quantum number $\ell $ for which the potential $V(r)$
possesses bound states (entailing of course that for $\ell \leq L_{\text{%
NLL3L}}^{(-)}$ the potential $V(r)$ does certainly possess at least one $%
\ell $-wave bound state): 
\end{subequations}
\begin{equation}
\text{NLL3L:\quad }L\geq L_{\text{NLL3L}}^{(-)}=\left\{ \left\{ \pi \left[
\log \left( \frac{q}{p}\right) \right] ^{-1}\nu \right\} \right\} ,
\label{NLL2L}
\end{equation}%
of course with $p,q$ and $\nu $ defined by (\ref{P}), (\ref{Q}) and (\ref{nu}%
). Here of course the double braces denote the integer part.

\subsection{Limits defined in terms of local properties of the potential
(not involving integrals over the potential)}

\label{sec1.3}

In this subsection we report a \textit{new }lower limit on the number of $%
\ell $-wave bound states $N_{\ell },$ which depends on the potential only
via the quantity $\sigma $, see (\ref{sigma}). Note that, perhaps with a
slight abuse of language, we consider (see the title of this section) the
quantity $\sigma $ to depend only on local properties of the potential,
since to calculate it only the value(s) of $r$ at which the function $%
2V(r)+r\,V^{\prime }(r)$ vanishes must be identified. We also provide, in
terms of this quantity $\sigma $, \textit{new} upper and lower limits on the
largest value $L$ of $\ell $ for which the potential $V(r)$ possesses bound
states.

The lower limit on $N_{\ell }$, which we denote NLL4, holds provided the
potential $V(r)$ satisfies, for all values of $r,$ $0\leq r<\infty $, the
inequality (\ref{CondV1}), which as we already noted above is automatically
satisfied by monotonically nondecreasing potentials, see (\ref{V1}). It
takes the neat form

\begin{equation}
\text{NLL4:\quad }N_{\ell }>-\frac{1}{2}+\frac{\sigma }{2(2\ell +1)}.
\label{NLL3}
\end{equation}
This lower limit features the correct power growth, see (\ref{NsubLasy1}),
as $g$ (see (\ref{gDef})) diverges, and it is best possible, being saturated
by the potential (\ref{VeqR4}) with (\ref{Alpha1}). The analogy of this
lower limit NLL4, see (\ref{NLL3}), with the lower limit C$\ell $n, see (\ref%
{Cl}), is remarkable; note that, since obviously $\sigma \geq (2/\pi )\rho
\left\vert V(\rho )\right\vert ^{1/2}$ (see (\ref{sigma})), this \textit{new}
limit, NLL4, would always be more stringent than C$\ell$n, were it not for
the additional factor $1/2$ multiplying $\sigma $ in the right-hand side of
the inequality (\ref{NLL3}) (in comparison to (\ref{Cla})).

This result, (\ref{NLL3}), clearly entails the following \textit{new} lower
limit $L_{\text{NLL4L}}^{(-)}$ on the largest value $L$ of $\ell $ for which
the potential $V(r)$ possesses bound states (entailing of course that for $%
\ell \le L_{\text{NLL4L}}^{(-)}$ the potential $V(r)$ does certainly possess
at least one $\ell $-wave bound state): 
\begin{equation}
\text{NLL4L:\quad }L\geq L_{\text{NLL4L}}^{(-)}=\left\{ \left\{ \frac{1}{2}%
(\sigma -1)\right\}\right\} .  \label{NLL3L}
\end{equation}
Note that this lower limit features as well the correct power growth, see (%
\ref{Lasy}), as $g$ (see (\ref{gDef})) diverges, and is best possible, being
saturated by the potential (\ref{VeqR4}) with (\ref{Alpha1}).

Let us recall that a somewhat analogous upper limit $L_{\text{eff}}^{(+)}$
on the largest value $L$ of $\ell $ for which the potential $V(r)$ possesses
bound states (entailing of course that for $\ell >L_{\text{eff}}^{(+)}$ the
potential $V(r)$ certainly does not possess any $\ell $-wave bound state),
which we denote as ULL, reads 
\begin{equation}
\text{ULL:}\quad L\leq L_{\text{eff}}^{(+)}=\left\{ \left\{ \frac{1}{2}(\pi
\sigma -1)\right\} \right\} .  \label{ULL}
\end{equation}%
[Indeed, it is an immediate consequence -- via a standard comparison
argument, see below -- of the well-known fact that the solution $u(r)$
characterized by the boundary condition $u(0)=0$ of the ODE $%
r^{2}\,u^{\prime \prime }(r)+c\,u(r)=0$ features a zero in $0<r<\infty $
only if the real constant $c$ exceeds $\frac{1}{4},$ $c>\frac{1}{4}$].

\subsection{Limits defined in terms of comparison potentials}

\label{sec1.4}

The results reported in this subsection are directly based on the elementary
remark that, if $V^{(1)}(r)\leq V^{(2)}(r)$ for all values of $r,$ $0\leq
r<\infty ,$ then the number $N_{\ell }^{(2)}$ of $\ell $-wave bound states
associated with the potential $V^{(2)}(r)$ cannot exceed the number $N_{\ell
}^{(1)}$ of $\ell $-wave bound states associated with the potential $%
V^{(1)}(r),$ $N_{\ell }^{(1)}\geq N_{\ell }^{(2)}.$

Let $V(r)$ satisfy the negativity condition (\ref{V2}) and let $H_{\lambda
}^{(\ell )}(r)$ be the \textquotedblleft additional\textquotedblright\ ($%
\ell $-dependent) potential defined as follows (see Section \ref{sec3}):

\begin{equation}
H_{\lambda }^{(\ell )}(r)=-\frac{\ell (\ell +1)}{r^{2}}+\frac{5}{16}\left(%
\frac{V^{\prime }(r)}{V(r)}\right) ^{2}+\frac{V^{\prime \prime }(r)}{%
4\left\vert V(r)\right\vert }+(1-4\lambda ^{2})\left\vert \,V(r)\right\vert,
\label{DefH}
\end{equation}
with $\lambda $ an arbitrary \textit{nonnegative} constant, $\lambda \geq 0$%
. There holds then the following limits on the number $N_{\ell }$ of $\ell $%
-wave bound states possessed by the potential $V(r)$:

\begin{equation}
N_{\ell }\ge\left\{ \left\{ \lambda \,S\right\} \right\} \quad \text{if}%
\quad H_{\lambda }^{(\ell )}(r)\geq 0\quad \text{for}\quad 0\leq r<\infty,
\label{NHlow}
\end{equation}

\begin{equation}
N_{\ell }\le\left\{ \left\{ \lambda \,S\right\} \right\} \quad \text{if}%
\quad H_{\lambda }^{(\ell )}(r)\leq 0\quad \text{for}\quad 0\leq r<\infty,
\label{NHup}
\end{equation}
where of course $S$ is defined by (\ref{S}) and the double braces denote the
integer part. Note however that, for higher partial waves ($\ell >0$), the
lower limit, (\ref{NHlow}), is applicable only to potentials that vanish at
the origin ($r=0$) at least proportionally to $r^{4\ell }$ and
asymptotically ($r\rightarrow \infty $) no faster than $r^{-4(\ell +1)}$;
while for S-waves ($\ell =0$), the upper limit is only applicable to
potentials that vanish asymptotically proportionally to $r^{-p}$ with (see (%
\ref{VatInf})) $2<p\leq 4$.

Note in particular that (the special case with $\ell =0$ and $\lambda =1/2$
of) this result implies that, for any potential $V(r)$ that satisfies, in
addition to the negativity condition (\ref{V2}), the inequality 
\begin{equation}
\frac{5}{4}\left( \frac{V^{\prime }(r)}{V(r)}\right) ^{2}-\frac{V^{\prime
\prime }(r)}{V(r)}\geq 0\quad \text{for\quad }0\leq r<\infty ,
\end{equation}%
there holds the following \textit{new} lower bound on the number $N_{0}$ of
S-wave bound states: 
\begin{equation}
N_{0}\geq \left\{ \left\{ \frac{S}{2}\right\} \right\} .
\end{equation}%
As can be easily verified, this lower limit is for instance applicable to
the (class of) potential(s) 
\begin{equation}
V(r)=-\frac{g^{2}}{R^{2}}\,\left( \frac{r}{R}\right) ^{\alpha -2}\,\exp %
\left[ -\left( \frac{r}{R}\right) ^{\beta }\right]
\end{equation}%
where $\alpha $ and $\beta $ are two arbitrary \textit{positive} constants, $%
\alpha >0$, $\beta >0$, that satisfy the following condition: 
\begin{equation}
\alpha \beta \geq \beta ^{2}+1.  \label{condit1}
\end{equation}%
It then yields the explicit lower limit 
\begin{equation}
N_{0}\geq \left\{ \left\{ \frac{g}{\pi \beta }\,2^{\frac{\alpha }{2\beta }%
}\,\Gamma \left( \frac{\alpha }{2\beta }\right) \right\} \right\} .
\end{equation}%
In particular, when $\alpha =2$ and $\beta =1$, we obtain the lower limit $%
N_{0}\geq \{\{2g/\pi \}\}$ on the number of S-wave bound states for the
exponential potential which simplifies and improves the lower bound found in
our previous work (see eq. (2.13) of Ref. \cite{BC}).

The particular case of the special potential $V(r)$ that yields via (\ref%
{DefH}) $H_{\lambda }^{(\ell )}(r)=0$ is investigated elsewhere \cite{BC2}.

\subsection{Limits of second kind, defined in terms of recursive formulas}

\label{sec1.5}

In this section we exhibit \textit{new }upper and lower limits on the number
of bound states, defined in terms of recursive formulas which are
particularly convenient for numerical computation. We call these limits
\textquotedblleft of second kind,\textquotedblright\ following the
terminology introduced in Ref. \cite{BC}. It is possible, following \cite{BC}%
, to derive such limits directly for the number $N_{\ell }$ of $\ell $-wave
bound states possessed by a central potential $V(r)$ having some
monotonicity properties, via a treatment based on the ODE (\ref{EqEta1})
(see Section \ref{sec3}) and utilizing the potential (\ref{VeqR4}) that, as
discussed in Section \ref{sec3}, trivializes the solution of this ODE (just
as the square-well potential employed in Ref. \cite{BC} to obtain this kind
of results trivializes the S-wave version of this ODE). But the results we
obtained in this manner, including the precise monotonicity conditions on
the potential $V(r)$ required for their validity (although, as a matter of
fact, the simple monotonicity condition (\ref{V1}) would be more than enough
for the validity of the upper limit), are not sufficiently neat, nor are
they expected to be sufficiently stringent, to warrant our reporting them
here. We rather focus on the derivation, using essentially the same
technique employed in Ref. \cite{BC}, of \textit{new }upper and lower limits
on the number $N_{0}$ of S-wave bound states possessed by a (nonmonotonic)
potential $V(r)$ that has two zeros, $V(r_{\pm })=0,$ and that is positive
for $r$ smaller than $r_{-},$ negative for$\ r$ in the interval from $r_{-}$
to $r_{+}$ with only one minimum, say at $r=r_{\text{min}},$ in this
interval, and is again positive for $r$ larger than r$_{+},$ see (\ref{V+-}%
). Indeed, as already noted in subsection \ref{sec1.2}, one can then obtain 
\textit{new} upper and lower limits on the number $N_{\ell }$ of $\ell $%
-wave bound states by replacing the potential $V(r)$ with the effective
potential $V_{\ell ,\text{eff}}(r)$, see (\ref{Veff}), since such a
potential, for a fairly large class of potentials $V(r)$, does indeed
satisfy the shape conditions mentioned above: see (\ref{V+-l}).

To get the \textit{new }upper limit one introduces the following two
recursions: 
\begin{equation}
r_{j+1}^{(\text{up,incr)}}=r_{j}^{(\text{up,incr)}}+\frac{\pi }{2\,}%
\,\left\vert V(r_{j}^{(\text{up,incr)}})\right\vert ^{-1/2},\quad \text{with}%
\quad r_{0}^{(\text{up,incr)}}=r_{\text{min}},
\end{equation}%
\begin{equation}
r_{j+1}^{(\text{up,decr)}}=r_{j}^{(\text{up,decr)}}-\frac{\pi }{2\,}%
\,\left\vert V(r_{j}^{(\text{up,decr)}})\right\vert ^{-1/2},\quad \text{with}%
\quad r_{0}^{(\text{up,decr)}}=r_{\text{min}},  \label{RecUpDecr}
\end{equation}%
that define the increasing respectively decreasing sequences of radii $%
r_{j}^{(\text{up,incr)}}$ respectively $r_{j}^{(\text{up,decr)}},$ both
starting from the value $r_{\text{min}}$ at which the potential attains its
minimum value, see (\ref{V+-}). Now let $J^{(\text{up,incr})}$ be the first
value of $j$ such that $r_{j}^{(\text{up,incr)}}$ exceeds or equals $r_{+}$, 
\begin{equation}
r_{J^{(\text{up,incr)}}-1}^{(\text{up,incr)}}<r_{+}\leq r_{J^{(\text{up,incr)%
}}}^{(\text{up,incr)}},  \label{Jupincr}
\end{equation}%
and likewise let $J^{(\text{up,decr})}$ be the first value of $j$ such that $%
r_{j}^{(\text{up,decr)}}$ becomes smaller than, or equal to, $r_{-}$, 
\begin{equation}
r_{J^{(\text{up,decr)}}}^{(\text{up,decr)}}\leq r_{-}<r_{J^{(\text{up,decr)}%
}-1}^{(\text{up,decr)}}.  \label{Jupdecr}
\end{equation}%
The \textit{new} upper limit of the second kind (ULSK) is then provided by
the neat formula 
\begin{equation}
\text{ULSK}:\quad N_{0}<\frac{1}{2}\left[ J^{(\text{up,incr})}+J^{(\text{%
up,decr})}+1+\theta \left( r_{J^{(\text{up,decr)}}}^{(\text{up,decr)}%
}\right) \right] .  \label{UpLiSecKind}
\end{equation}%
Here $\theta (x)$ is the standard step function, $\theta (x)=1$ if $x\geq 0$%
, $\theta (x)=0$ if $x<0$, and of course $r_{J^{(\text{up,decr)}}}^{(\text{%
up,decr)}}$ is the \textquotedblleft last\textquotedblright\ (smallest)
radius yielded by the recursion (\ref{RecUpDecr}), see (\ref{Jupdecr}). [Of
course the \textquotedblleft $\theta $-term\textquotedblright\ in the
right-hand side of this formula, (\ref{UpLiSecKind}), is not very
significant, at least for potentials possessing many bound states, namely
just when this upper limit is more likely to be quite cogent, see Section %
\ref{sec2}].

To obtain a lower limit one must instead define the following increasing
respectively decreasing sequences of radii $r_{j}^{(\text{lo,incr)}}$
respectively $r_{j}^{(\text{lo,decr)}}$: 
\begin{equation}
r_{j+1}^{(\text{lo,incr)}}=r_{j}^{(\text{lo,incr)}}+\frac{\pi }{2\,}%
\,\left\vert V(r_{j}^{(\text{lo,incr)}})\right\vert ^{-1/2},\quad \text{with}%
\quad r_{0}^{(\text{lo,incr)}}>r_{-},  \label{RecLoIncr}
\end{equation}%
\begin{equation}
r_{j+1}^{(\text{lo,decr)}}=r_{j}^{(\text{lo,decr)}}-\frac{\pi }{2\,}%
\,\left\vert V(r_{j}^{(\text{lo,decr)}})\right\vert ^{-1/2},\quad \text{with}%
\quad r_{0}^{(\text{lo,decr)}}<r_{+}.  \label{RecLoDecr}
\end{equation}%
Now let $J^{(\text{lo,incr})}$ be the first value of $j$ such that $r_{j}^{(%
\text{lo,incr)}}$ exceeds or equals $r_{\text{min}}$, 
\begin{equation}
r_{J^{(\text{lo,incr)}}-1}^{(\text{lo,incr)}}<r_{\text{min}}\leq r_{J^{(%
\text{lo,incr)}}}^{(\text{lo,incr)}},
\end{equation}%
and likewise let $J^{(\text{up,decr})}$ be the first value of $j$ such that $%
r_{j}^{(\text{up,decr)}}$ becomes smaller than, or equal to, $r_{\text{min}}$%
, 
\begin{equation}
r_{J^{(\text{lo,decr)}}}^{(\text{lo,decr)}}\leq r_{\text{min}}<r_{J^{(\text{%
lo,decr)}}-1}^{(\text{lo,decr)}}.
\end{equation}%
The \textit{new} lower limit of the second kind (LLSK) is then provided by
the neat formula 
\begin{equation}
\text{LLSK}:\quad N_{0}>\frac{1}{2}\left( J^{(\text{lo,incr})}+J^{(\text{%
lo,decr})}-H\right) -1,  \label{LoLiSecKind}
\end{equation}%
where the parameter $H$ vanishes, $H=0,$ provided either $\left\vert V\left(
r_{J^{(\text{lo,incr)}}-1}^{(\text{lo,incr)}}\right) \right\vert \leq $ $%
\left\vert V\left( r_{J^{(\text{lo,decr)}}-1}^{(\text{lo,decr)}}\right)
\right\vert $ and $r_{J^{(\text{lo,incr)}}}^{(\text{lo,incr)}}$ does not
exceed $r_{J^{(\text{lo,decr)}}-1}^{(\text{lo,decr)}}$, $r_{J^{(\text{%
lo,incr)}}}^{(\text{lo,incr)}}\leq r_{J^{(\text{lo,decr)}}-1}^{(\text{%
lo,decr)}},$ or $\left\vert V\left( r_{J^{(\text{lo,incr)}}-1}^{(\text{%
lo,incr)}}\right) \right\vert \geq $ $\left\vert V\left( r_{J^{(\text{%
lo,decr)}}-1}^{(\text{lo,decr)}}\right) \right\vert $ and $r_{J^{(\text{%
lo,incr)}}-1}^{(\text{lo,incr)}}$ does not exceed $r_{J^{(\text{lo,decr)}%
}}^{(\text{lo,decr)}}$, $r_{J^{(\text{lo,incr)}}-1}^{(\text{lo,incr)}}\leq
r_{J^{(\text{lo,decr)}}}^{(\text{lo,decr)}}$, and it is unity, $H=1$,
otherwise. [Anyway this term does not make a very significant contribution,
at least for potentials possessing many bound states, when this lower limit
is more likely to be quite cogent, see Section \ref{sec2}]. Note moreover
that, in the recursions (\ref{RecLoIncr}) respectively (\ref{RecLoDecr}),
the starting points, $r_{0}^{(\text{lo,incr)}}$ respectively $r_{0}^{(\text{%
lo,decr)}},$ are only restricted by inequalities; of course interesting
results will obtain only by assigning $r_{0}^{(\text{lo,incr)}}$ relatively,
but not exceedingly, close to $r_{-}$, and $r_{0}^{(\text{lo,decr)}}$
relatively, but not exceedingly, close to $r_{+}$ [to get some understanding
of which choices of these parameters, $r_{0}^{(\text{lo,incr)}}$ and $%
r_{0}^{(\text{lo,decr)}},$ are likely to produce more cogent results, the
interested reader is referred to the proof of the lower limit given in
Section \ref{sec3}; of course numerically one can make a search for the
values of these parameters, $r_{0}^{(\text{lo,incr)}}$ respectively $r_{0}^{(%
\text{lo,decr)}},$ that maximize the right-hand side of (\ref{LoLiSecKind}),
starting from values close to $r_{-}$ respectively $r_{+}$].

\subsection{Limits on the total number of bound states}

\label{sec1.6}

Clearly if $N_{\ell }^{(-)}$ respectively $N_{\ell }^{(+)}$ provide lower,
respectively upper, limits on the number $N_{\ell }$ of $\ell $-wave bound
states, and likewise $L^{(-)},$ respectively $L^{(+)},$ provide lower,
respectively upper, limits on the largest value $L$ of the angular momentum
quantum number $\ell $ for which the potential $V(r)$ does possess bound
states, it is plain that the quantities 
\begin{subequations}
\label{Nuplo}
\begin{equation}
N^{(\pm )}=N\left( L^{(\pm )}\right) ,  \label{Nuploa}
\end{equation}%
where 
\begin{equation}
N\left( L\right) =\sum_{\ell =0}^{L}\left( 2\ell +1\right) \,N_{\ell }^{(\pm
)},  \label{Nuplob}
\end{equation}%
provide lower respectively upper limits, 
\end{subequations}
\begin{equation}
N^{(-)}\leq N\leq N^{(+)},  \label{NNN}
\end{equation}%
to the total number $N$, see (\ref{TotNum}), of bound states possessed by
the potential $V(r).$ Hence several such limits can be easily obtained from
the results reported above.

\textit{Remark}. There is however a significant loss of accuracy in using
these formulas, (\ref{Nuplo}) and (\ref{NNN}), to obtain upper or lower
limits on the total number of bound states $N$. Note that of course the
upper limit, $L,$ of the sum in the right-hand side of (\ref{Nuplob}) must
be an integer, but after the sum has been performed to calculate $N(L),$ see
(\ref{Nuplob}), it gets generally replaced by a noninteger number, $L^{(+)}$
respectively $L^{(-)}$, to evaluate the upper respectively lower limit $%
N^{(+)}$ respectively $N^{(-)}$, see (\ref{Nuploa}) and (\ref{NNN}). Let us
illustrate this effect by a fictitious numerical example. Suppose we were
able to prove, say, the lower limit $N_{\ell }>\frac{13}{3}-\ell $. We would
then know that $N_{0}>\frac{13}{3}$, $N_{1}>\frac{10}{3}$, $N_{2}>\frac{7}{3}
$, $N_{3}>\frac{4}{3}$, $N_{4}>\frac{1}{3}$ entailing $N_{0}\geq 5$, $%
N_{1}\geq 4$, $N_{2}\geq 3$, $N_{3}\geq 2$, $N_{4}\geq 1$ hence we could
conclude that there are at least $55$ bound states ($5+12+15+14+9=55$), $%
N\geq 55$. But via the above procedure we would infer that $L^{(-)}=\frac{10%
}{3}$ and $N\left( L\right) =\frac{1}{6}\,\left( L+1\right) \,\left(
26+21L-4L^{2}\right) $ entailing $N^{(-)}>37.23$ hence $N\geq 38$. This is a
much less stringent (lower) limit. Clearly, due to the round off errors, a
lot of information got lost. This defect can be remedied, but only
marginally, by inserting in the expression $N(L)$ the best value of $L^{(-)}$
yielded by the above fictitious lower bound, namely $L^{(-)}=4,$ since we
then obtain $N^{(-)}>38.3$ hence $N\geq 39$. And the analogous calculation
via (\ref{Nuplo}) and (\ref{NNN}) from an hypothetical upper limit $N_{\ell
}<\frac{13}{3}-\ell $, which clearly entails $N_{0}\leq 4$, $N_{1}\leq 3$, $%
N_{2}\leq 2$, $N_{3}\leq 1$ and hence $N\leq 30$, yields again a less
stringent result, namely the upper limit $N\leq 37$ if $L^{(+)}=\frac{10}{3}$
is used. This limit can be slightly improved, namely $N\leq 35$, if the
integer part of $L^{(+)}=3$ is used. In the following we have tried to take
care of this problem -- to the extent possible compatibly with the goal to
obtain simple explicit formulas.

In the next section we illustrate the remark just made by computing firstly,
via the upper and lower limits NUL2$\ell $ and NLL2$\ell ,$ two sets of
integers $N_{\ell }^{(-)}$ and $N_{\ell }^{(+)}$ such that $N_{\ell
}^{(-)}\leq N_{\ell }\leq N_{\ell }^{(+)},$ and by then evaluating upper and
lower limits $N^{(\pm )}$ on the total number $N$ of bound states, see (\ref%
{NNN}), via the standard formula (\ref{Nuplob}) with $L$ replaced, as it
were, by $\infty $, the sum being automatically stopped by the vanishing of
the summand. The upper and lower limits obtained with this procedure will be
called NUL2N and NLL2N respectively.

Anyway in this subsection some results obtained via (\ref{Nuplo}) and (\ref%
{NNN}) are reported, namely those we believe deserve to be displayed thanks
to their neat character. But firstly let us tersely review the upper limits
on the total number of bound states $N$ previously known (we did not find
any lower limits on $N$ in the literature).

A classical result, the validity of which is not restricted to central
potentials, is known in the literature as the Birman-Schwinger upper bound 
\cite{sch,bir}, and we denote it as BiS. It reads as follows: 
\begin{equation}
\text{BiS:\quad }N<\frac{1}{(4\pi )^{2}}\,\int d^{3}\vec{r}_{1}\,d^{3}\vec{r}%
_{2}\frac{V^{(-)}(\vec{r}_{1})\,V^{(-)}(\vec{r}_{2})}{\left\vert \vec{r}_{1}-%
\vec{r}_{2}\right\vert ^{2}},  \label{BiS}
\end{equation}
implying, for central potentials, 
\begin{equation}
\text{BiScentral:\quad }N<\frac{1}{2}\int_{0}^{\infty}dr_{1}\,r_{1}%
\,V^{(-)}(r_{1})\,\int_{0}^{\infty }dr_{2}\,r_{2}\,V^{(-)}(r_{2})\,\log
\left\vert \frac{r_{1}+r_{2}}{r_{1}-r_{2}}\right\vert .
\end{equation}
This upper limit, however, is proportional to $g^{4}$ (see (\ref{gDef}))
rather than $g^{3}$ (see (\ref{NpropG3})), hence it provides a limit much
larger than the exact result for strong potentials possessing many bound
states.

A simple upper limit, that we denote BSN, can be obtained from the BS$\ell $
upper limit, see (\ref{BS1}); it reads

\begin{equation}
\text{BSN:\quad }N<\left\{ \left\{ I\right\} \right\} \left\{ \left\{ \frac{%
I+1}{2}\right\} \right\} ,  \label{BSN}
\end{equation}
with

\begin{equation}
I=\int_{0}^{\infty }dr\,r\,\left[- V^{(-)}(r)\right].
\end{equation}
This upper limit is also proportional to $g^{4}$ rather than $g^{3}$.

An upper limit that does not have this defect and that is also valid for
potentials that need not be central was obtained by E. Lieb \cite{L}. We
denote it as L: 
\begin{equation}
\text{L:\quad }N<0.116\int d^{3}\vec{r}\,\left[- V^{(-)}(\vec{r}\,)\right]^{3/2}  \label{L}
\end{equation}%
(for the origin of the numerical coefficient in the right-hand side of this
formula, we refer to the original paper \cite{L}). For central potentials it
reads as follows: 
\begin{equation}
\text{Lcentral:\quad }N<1.458\int_{0}^{\infty }dr\,r^{2}\,\left[-V^{(-)}(r)\right]^{3/2}  \label{Lc}
\end{equation}%
(the numerical coefficient in this formula is of course obtained by
multiplying that in the preceding formula by $4\pi $; for other results of
this kind, none of which seems however to be more stringent than those
reported here, see \cite{BlSt}).

Let us end this listing of previously \textit{known} results by reporting
the upper limit on the total number of bound states $N$ obtained \cite{CMS}
by inserting (\ref{CMSn}) and (\ref{CMSL}) in (\ref{Nuplo}). As entailed by
its origin, it only holds for monotonically nondecreasing potentials, see (%
\ref{V1}) and (\ref{V2}). We denote it as CMSN: 
\begin{equation}
\text{CMSN:\quad }N<\frac{\pi ^{2}}{12}\,\left[ S^{3}+3S^{2}+\frac{2}{\pi }
\left( 3-\frac{1}{2\pi }\right) S+\frac{3}{\pi ^{2}}\right] ,  \label{CMSN}
\end{equation}
with $S$ defined by (\ref{S}).

We did not obtain any \textit{new} upper limit on the total number $N$ of
bound states sufficiently neat to be worth reporting. We report instead a
rather trivial\ upper limit on $N$ obtained via (\ref{Nuplo}) with $L$
replaced by its upper limit $L_{\text{eff}}^{(+)}$ (see (\ref{ULL})) and
with $N_{\ell }\leq N_{0}$ and $N_{0}$ bounded above by NUL2, see (\ref%
{NUL2bis}). This upper limit on the total number $N$ of bound states is
therefore applicable to potentials that satisfy the condition (\ref{Vr+-}),
and we denote it as NUL2Nn: 
\begin{equation}
\text{NUL2Nn}:\quad N<\frac{1}{8}(\pi \sigma +1)^{2}\,\left\{ 2+S+\frac{1}{%
\pi }\log \left[ \frac{-V^{(-)}(r_{\text{min}})}{M}\right] \right\} ,
\label{NUL2N}
\end{equation}%
with $M$ defined by (\ref{Mdef}), $\sigma $ defined by (\ref{sigma}), $S$
defined by (\ref{S}) and $V^{(-)}(r_{\text{min}})$ the minimal value of (the
negative part of) the potential, see (\ref{Vr+-}). For a \textit{monotonic }%
potential, see (\ref{V1}), this upper limit takes the simpler form%
\begin{equation}
\text{NUL2Nm}:\quad N<\frac{1}{8}(\pi \sigma +1)^{2}\,\left\{ 1+S+\frac{1}{%
2\pi }\log \left[ \frac{V(p)}{V(q)}\right] \right\} ,  \label{NUL2Nb}
\end{equation}%
where $p$ and $q$ are defined by (\ref{P}) and (\ref{Q}) respectively (this result is of course obtained using, instead of NUL2, the analogous result valid for monotonic potentials \cite{BC}). It is
remarkable that, in spite of the drastic approximation $N_{\ell }\leq N_{0}$
used to get these two limits, they turn out, in all the tests performed in
Section \ref{sec2}, to be more stringent than all previously known results.

We now report two \textit{new} lower limits, which recommend themselves
because of their neatness, although, for the reason outlined above, one
cannot expect them to be very stringent.

A \textit{new} lower limit, that we denote as NLLN3, on the total number $N$
of bound states for a potential $V(r)$ that satisfies the conditions (\ref%
{Vr+-}), follows from the lower limits NLL3, see (\ref{NLL2}), and NLL3L,
see (\ref{NLL2L}). A simple calculation yields 
\begin{equation}
\text{NLLN3:\quad }N>\frac{\nu}{6\lambda^2}(2\nu+\lambda)(\nu+\lambda),
\label{NLLN2}
\end{equation}
with $\nu $ defined by (\ref{nu}) and 
\begin{equation}
\lambda =\frac{1}{\pi }\log \left( \frac{q}{p}\right)  \label{lambda}
\end{equation}
with $q$ and $p$ defined by (\ref{P}) and (\ref{Q}) (we assume of course $%
q\geq p$, hence $\lambda \geq 0$).

Another \textit{new }lower limit, that we denote as NLLN4, on the total
number $N$ of bound states for a potential $V(r)$ that satisfies the
condition (\ref{CondV1}) is implied, via (\ref{Nuplo}), from the lower
limits NLL4, see (\ref{NLL3}), and NLL4L, see (\ref{NLL3L}). A simple
calculation yields 
\begin{equation}
\text{NLLN4:\quad }N\geq \frac{1}{2}\left\{ \left\{ \left( \frac{\sigma +1}{2%
}\right) \right\} \right\} \,\left\{ \left\{ \left( \frac{\sigma +3}{2}%
\right) \right\} \right\} ,  \label{NLL3N}
\end{equation}%
with $\sigma $ defined by (\ref{sigma}). Here, as usual, the double braces
denote the integer part. This lower limit has the merit of being rather
neat, but it grows proportionally to $g^{2}$ (see (\ref{gDef})) rather than $%
g^{3}$ (see (\ref{NpropG3})), hence it cannot be expected to be cogent for
strong potentials possessing many bound states.

\section{TESTS}

\label{sec2}

In this section we test the efficiency of the \textit{new} limits reported
in Section \ref{sec1} by comparing them for some representative potentials
with the exact results and with the results obtained via previously \textit{%
known} limits. For these tests we use three different potentials: the Morse
potential \cite{mors29} (hereafter referred to as M) 
\begin{equation}
\text{M:}\quad V(r)=-g^{2}\,R^{-2}\,\left\{ 2\exp \left[ -\left( \frac{r}{R}%
-\alpha \right) \right] -\exp \left[ -2\left( \frac{r}{R}-\alpha \right) %
\right] \right\} ;  \label{morse}
\end{equation}%
the exponential potential (hereafter referred to as E) 
\begin{equation}
\text{E:}\quad V(r)=-g^{2}\,R^{-2}\,\exp \left( -\frac{r}{R}\right) ;
\label{expo}
\end{equation}%
and the Yukawa potential (hereafter referred to as Y) 
\begin{equation}
\text{Y:}\quad V(r)=-g^{2}\,(rR)^{-1}\,\exp \left( -\frac{r}{R}\right) .
\label{yuk}
\end{equation}%
In all these equations, and below, $R$ is an arbitrary (of course positive)
given radius, and $g$, as well as $\alpha $ in (\ref{morse}), are arbitrary
dimensionless \textit{positive }constants.

\subsection{Tests of the limits on the number of bound states $N_{\ell}$}

\label{sec2.1}

The first potential we use to test the new limits is the M potential (\ref%
{morse}). This is a nonmonotonic potential for which the number $N_{0}$ of
bound states for vanishing angular momentum is known; we indeed consider for
this potential only the $\ell =0$ case. [We do not test the GGMT and CMS2
limits with this M potential since, from their incorrect behavior when the
strength $g$ of the potential diverges, we already know that these limits
give poor results. But, in spite of this incorrect behavior, these limits
could be useful when there are few bound states; they are therefore tested
below with the E and Y potentials, in cases with $\ell >0$].

The exact formula for the number of S-wave bound states for the M potential
is 
\begin{equation}
N_{0}=\left\{ \left\{ g+\frac{1}{2}\right\} \right\} .  \label{morse_ex}
\end{equation}%
Note that it is independent of the value of the constant $\alpha $.

For this potential, the limits NUL2 and NLL2, see (\ref{NUL2bis}) and (\ref%
{NLL2bis}), can be computed (almost completely) analytically:%
\begin{eqnarray}
\text{NUL2}:\quad N_{0} &<&g-\frac{1}{2\pi }\log s+1,  \label{nl2_morse} \\
\text{NLL2}:\quad N_{0} &>&g+\frac{1}{2\pi }\log s-\frac{3}{2},
\end{eqnarray}%
with $s=\min (2y-y^{2},2x-x^{2})$ and $x$, $y$ solutions of 
\begin{eqnarray}
\pi -\sqrt{2y-y^{2}}-2\arcsin \left( \frac{y}{2}\right) &=&\frac{\pi }{2g},
\label{sol_morse} \\
\sqrt{2x-x^{2}}+2\arcsin \left( \frac{x}{2}\right) &=&\frac{\pi }{2g}
\end{eqnarray}%
The calculation of the cutoff radii $p$ and $q$, see (\ref{P}) and (\ref{Q}%
), cannot be evaluated analytically. But one can compute upper and lower
limits, $\tilde{p}<p$ and $\tilde{q}>q$, on these radii by using only the
attractive part of the potential in the definition (\ref{P}) and (\ref{Q})
of $p$ and $q$. When $\tilde{p}$ and $\tilde{q}$ are used in place of $p$
and $q$ we obtain the (marginally less stringent) limits (denotes as NUL2s
and NLL2s) 
\begin{eqnarray}
\text{NUL2s}:\quad N_{0} &<&g+\frac{1}{2\pi }\log \left[ \frac{z^{4}}{%
4(z^{2}-1)}\right] +1,  \label{nl2b_morse} \\
\text{NLL2s}:\quad N_{0} &>&g-\frac{1}{2\pi }\log \left[ \frac{z^{4}}{%
4(z^{2}-1)}\right] -\frac{3}{2},  \label{nl2b_morse2}
\end{eqnarray}%
with $z=8g/\pi $. As mentioned in Section \ref{sec1}, validity of the
inequalities $\tilde{p}\leq r_{\text{min}}\leq \tilde{q}$ is required in
order to use the NUL2s and NLL2s limits; this leads to the restriction $%
g\geq \pi \sqrt{2}/(8(\sqrt{2}-1))\cong 1.34$.

The NLL1 limit (\ref{NLL1}) takes for the M potential the simple form 
\begin{equation}
\text{NLL1:}\quad N_{0}>0.672\,g-1.  \label{NLL1M}
\end{equation}

\begin{figure}
\begin{center}
\includegraphics[width=10cm]{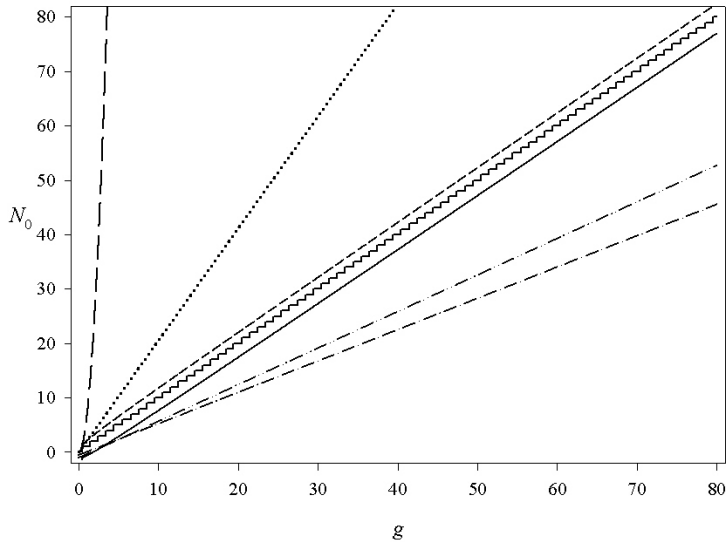}
\end{center}
\caption{Comparison between the exact value (\protect\ref{morse_ex}) of $%
N_{0}$ (ladder), the upper limits BS$\ell$ (\protect\ref{BSlm}) (long dash),
M$\ell$ (\protect\ref{Mlm}) (spare dot) and NUL2s (\protect\ref{nl2b_morse})
(short dash) and the lower limits C$\ell$ (\protect\ref{Clm}) (dash-dot),
NLL2s (\protect\ref{nl2b_morse2}) (solid) and NLL1 (\protect\ref{NLL1M})
(dash-dot-dot) for the M potential (\protect\ref{morse}) (all with $\ell =0$ and $\protect\alpha =1
$).}
\label{fig1}
\end{figure}

The other \textit{new} limits cannot be tested with this potential: the
upper limit NUL1 and the limits of the second kind are not applicable
because $r_{+}=\infty $, the lower limit NLL4 is only applicable to monotone
potentials, and the lower limit NLL3 coincides with NLL2 for $\ell =0$.

The previously known limits applicable to this potential (note that the CMS
upper limit is only applicable to monotone potentials) take the form: 
\begin{equation}
\text{BS}\ell :\quad N_{0}\leq 2g^{2}(\alpha +\log 2+\frac{3}{2});
\label{BSlm}
\end{equation}%
\begin{equation}
\text{M}\ell :\quad N_{0}\leq \sqrt{2}g(\alpha ^{2}+(3-2\log 2)\alpha
+1.901)^{1/4};  \label{Mlm}
\end{equation}%
\begin{subequations}
\label{Clm}
\begin{eqnarray}
\label{Clma}
\text{C}\ell :\quad N_{0} &>&-\frac{1}{2}+\frac{2}{\pi }g\,T(\alpha ),
 \\
\label{Clmb}
T(\alpha ) &=&\max_{0\leq \gamma \leq 1}\left\{ \frac{1}{\gamma }\left[ \exp
(\alpha )-\frac{1}{4}\exp (2\alpha )-\sqrt{1-\gamma ^{2}}+\frac{\gamma ^{2}}{%
2}\log \left( \frac{1+\sqrt{1-\gamma ^{2}}}{1-\sqrt{1-\gamma ^{2}}}\right) %
\right] \right\} .
\end{eqnarray}%
To obtain this limit we set $a=R/(\gamma g)$ in (\ref{loCal}) with $\ell =0$%
. Note that for $\alpha >\alpha _{0}\cong 1.386$, with 4$\,\exp (\alpha
_{0})=\exp (2\alpha _{0})$, this lower limit C$\ell $ is trivial because $%
T(\alpha )$ is then negative). Note that, for $\alpha =\log 2,$ $T(\alpha )$
reach is maximal value: $T(\alpha )=1.055$. The factor multiplying $g$ in
the right-hand side of (\ref{Clma}) coincides which that multiplying $g$ in
the right-hand side of the NLL1 lower limit (\ref{NLL1M}); so for this
particular value of $\alpha $, this C$\ell $ limit is slightly more
stringent (of course for all values of $g)$ than the NLL1 lower limit. For
all other values of $\alpha $, there exists a value $g_{\alpha }$ such as
for all $g\geq g_{\alpha }$, the NLL1 limit is more stringent than the C$%
\ell $ limit. For example, for $\alpha =1$, the NLL1 limit yield more cogent
results than the C$\ell $ limit as soon as $g\geq 5.22$, namely as soon as
the number of bound states is greater than five.

Fig. \ref{fig1} displays these limits as a function of $g$. Note that the
limits depend on $\alpha $ while the exact result does not. We tested the
results for the $\alpha =1$ case (not $\alpha =0,$ in oder to have a
nonmonotonic potential). It is clear from this figure that the
generalizations to nonmonotonic potentials of the results obtained in Ref. 
\cite{BC} , namely the limits NUL2 and NLL2, are quite cogent. This remains
true even for large values of $g:$ for instance, when the exact number $%
N_{0} $ of bound states is 5000, these upper and lower limits restrict its
value to the rather small interval $[4996,5003]$. In this case the BS$\ell $
upper limit exceeds $1.5\ 10^{8}$, the M$\ell $ upper limit only informs us
that $N<10307$, the lower limit C$\ell $ that $N>2879$ and the lower limit
NLL1 that $N>3359$.

The second test is performed with the E potential (\ref{expo}). The exact
number $N_{\ell }$ of bound states for this potential is computed by
integrating numerically (\ref{EqEta1}) with $\eta (0)=0$ and $\eta (\infty
)=N_{\ell }\pi $, see Section \ref{sec3}.

The upper limit NUL1$\ell $ reads 
\end{subequations}
\begin{equation}
\text{NUL1}\ell :\quad N_{\ell }<1+\frac{2}{\pi }\sqrt{x_{+}-x_{-}}\left\{
g^{2}\,\left[ \exp (-x_{-})-\exp (-x_{+})\right] -\ell (\ell +1)\left( \frac{%
1}{x_{-}}-\frac{1}{x_{+}}\right) \right\} ^{1/2},  \label{nul1_exp}
\end{equation}%
where $x_{\pm }$ are the two solutions of 
\begin{equation}
\ell (\ell +1)=g^{2}\,x_{\pm }^{2}\,\exp (-x_{\pm }).  \label{xpm}
\end{equation}

The NLL1n$\ell $ limit reads 
\begin{equation}
\text{NUL1n}\ell :\quad N_{\ell }>-1+\frac{1}{\pi }\left\{ g^{2}\left[ \exp
(-x_{-})-\exp (-x_{+})\right] -\ell (\ell +1)\left( \frac{1}{x_{-}}-\frac{1}{%
x_{+}}\right) \right\} \ |V_{\ell ,\text{eff}}^{\text{min}}|^{-1/2},
\label{nll1_exp}
\end{equation}%
where $V_{\ell ,\text{eff}}^{\text{min}}$ is the minimal value of the
effective potential (\ref{Veff}). The NUL2$\ell $ and NLL2$\ell $ limits can
be written as follows: 
\begin{eqnarray}
\text{NUL2}\ell :\quad N_{\ell } &<&\frac{1}{\pi }F(g,\ell ;x_{-},x_{+})+%
\frac{1}{2\pi }\log \left\vert \frac{V_{\ell ,\text{eff}}^{\text{min}}}{M}%
\right\vert +1,  \label{nul2_exp} \\
\text{NLL2}\ell :\quad N_{\ell } &>&\frac{1}{\pi }F(g,\ell ;x_{-},x_{+})-%
\frac{1}{2\pi }\log \left\vert \frac{V_{\ell ,\text{eff}}^{\text{min}}}{M}%
\right\vert -\frac{3}{2},
\end{eqnarray}%
where 
\begin{equation}
F(g,\ell ;a,b)=\int_{a}^{b}\frac{dx}{x}\sqrt{g^{2}x^{2}\exp (-x)-\ell (\ell
+1)},  \label{fab}
\end{equation}%
and 
\begin{subequations}
\begin{equation}
M=\min \left[ |V_{\ell ,\text{eff}}(p)|,|V_{\ell ,\text{eff}}(q)|\right] 
\end{equation}%
where $p$ and $q$ are solutions of 
\begin{equation}
F(g,\ell ;x_{-},\frac{p}{R})=\frac{\pi }{2},\quad F(g,\ell ;\frac{q}{R}%
,x_{+})=\frac{\pi }{2}.
\end{equation}

The lower limits NLL3 and NLL4 take much simpler forms: 
\end{subequations}
\begin{equation}
\text{NLL3}:\quad N_{\ell }>\frac{2}{\pi }\,g-\frac{1}{2\pi }\log \left( 
\frac{4g}{\pi }\right) -\frac{3}{2}-\frac{\ell }{\pi }\log \left[ \frac{\log
x}{\log (1-x)}\right] ,  \label{nul3_exp}
\end{equation}%
with $x=\pi /(4g)$, and 
\begin{equation}
\text{NLL4}:\quad N_{\ell }>\frac{2g}{\pi e(2\ell +1)}-\frac{1}{2}.
\label{nul4_exp}
\end{equation}%
The previously known limits are found to be: 
\begin{equation}
\text{BS}\ell :\quad N_{\ell }<\frac{g^{2}}{2\ell +1};
\end{equation}%
\begin{equation}
\text{CMS}:\quad N_{\ell }<\frac{4g}{\pi }+1-\sqrt{1+\frac{4\,\ell \,(\ell
+1)}{\pi ^{2}}};
\end{equation}%
\begin{subequations}
\begin{equation}
\text{M}\ell :\quad N_{\ell }<(AB)^{1/4}
\end{equation}%
where $A$ and $B$ are given by 
\begin{eqnarray}
A &=&g^{2}\,\left[ \exp (-x_{-})(x_{-}^{2}+2x_{-}+2)-\exp
(-x_{+})(x_{+}^{2}+2x_{+}+2)\right] -\ell (\ell +1)(x_{+}-x_{-}), \\
B &=&g^{2}\,\left[ \exp (-x_{-})-\exp (-x_{+})\right] -\ell (\ell +1)\left( 
\frac{1}{x_{-}}-\frac{1}{x_{+}}\right) ;
\end{eqnarray}%
\end{subequations}
\begin{equation}
\text{GGMT}:\quad N_{\ell }\leq g^{2p}(2\ell +1)^{(1-2p)}\,\frac{C_{p}\Gamma
(2p)}{p^{2p}},  \label{GGMTbis}
\end{equation}%
with $C_{p}$ defined by (\ref{GGMT2}); 
\begin{equation}
\text{CMS2}:\quad N_{\ell }<g^{2p}(2\ell +1)^{(1-2p)}\,\frac{\tilde{C}%
_{p}\Gamma (2p)}{p^{2p}},
\end{equation}%
with $\tilde{C}_{p}$ defined by (\ref{CMS2b}); 
\begin{equation}
\text{C}\ell :\quad N_{\ell }>\frac{2g}{\pi (2\ell +1)}\,y\exp \left(-\frac{y}{2}\right)-\frac{1}{2},
\end{equation}%
where $y$ is the solution of $y\exp (-y)=(2\ell +1)y^{2\ell }\Gamma (1-2\ell
,y)$.

\begin{table}[tbp]
\caption{Comparison for the E potential (\protect\ref{expo}) between the
exact number $N_{\ell}$ of bound states, various upper and lower limits on $%
N_{\ell}$ previously known and new upper and lower limits on $N_{\ell}$, for
several values of $g$ and $\ell$.}
\label{tab1}
\begin{center}
\begin{tabular}{cc|cccccccccccc}
$g$ & $\ell $ & LLSK & NLL3 & NLL1n$\ell$ & NLL2$\ell $ & Ex & NUL2$\ell $ & 
BS$\ell $ & CMS & M$\ell $ & GGMT & NUL1$\ell $ & ULSK \\ \hline
8 & 1 & 3 & 3 & 3 & 3 & \textbf{4} & 5 & 21 & 9 & 8 & 21 & 12 & 6 \\ 
& 3 & 1 & 1 & 2 & 1 & \textbf{2} & 3 & 9 & 8 & 5 & 6 & 6 & 4 \\ 
13 & 2 & 5 & 4 & 4 & 5 & \textbf{7} & 8 & 33 & 15 & 13 & 31 & 18 & 9 \\ 
& 6 & 2 & 0 & 2 & 2 & \textbf{3} & 4 & 13 & 13 & 7 & 7 & 7 & 5 \\ 
18 & 3 & 7 & 6 & 6 & 7 & \textbf{9} & 10 & 46 & 21 & 18 & 43 & 25 & 11 \\ 
& 9 & 2 & 0 & 3 & 2 & \textbf{4} & 4 & 17 & 17 & 9 & 8 & 9 & 5 \\ 
24 & 4 & 10 & 8 & 8 & 10 & \textbf{12} & 13 & 64 & 28 & 24 & 60 & 33 & 14 \\ 
& 12 & 4 & 0 & 4 & 4 & \textbf{5} & 6 & 23 & 23 & 13 & 11 & 12 & 7 \\ 
29 & 5 & 13 & 9 & 9 & 13 & \textbf{15} & 16 & 76 & 34 & 29 & 71 & 40 & 17 \\ 
& 15 & 4 & 0 & 5 & 4 & \textbf{6} & 7 & 27 & 28 & 15 & 13 & 13 & 8 \\ 
35 & 6 & 16 & 11 & 11 & 16 & \textbf{18} & 19 & 94 & 41 & 35 & 88 & 49 & 20
\\ 
& 18 & 6 & 0 & 6 & 6 & \textbf{7} & 8 & 33 & 33 & 18 & 16 & 16 & 9 \\ 
40 & 7 & 18 & 12 & 13 & 18 & \textbf{20} & 21 & 106 & 47 & 40 & 100 & 55 & 22
\\ 
& 21 & 6 & 0 & 6 & 6 & \textbf{8} & 9 & 37 & 38 & 20 & 18 & 18 & 10 \\ \hline
\end{tabular}%
\end{center}
\end{table}

Comparisons between the various limits and the exact results are presented
in Table \ref{tab1}. The BS$\ell $ limit gives poor results when $g$ becomes
large but becomes slightly better as $\ell $ grows. The CMS gives better
restrictions when $\ell $ is small but behaves like the BS$\ell $ limit when 
$\ell $ grows. The M$\ell $ limit overestimates the number of bound states
by a factor 2 when $\ell $ is small; it is no better for larger $\ell $, yet
better than the BS$\ell $ and CMS limits. The GGMT limit (with, in each
case, the optimized value of the parameter $p,$ see (\ref{GGMTbis})) gives
similar results to those yielded by the BS$\ell $ limit when $\ell $ is
small and becomes better and equivalent to the M$\ell $ limit for larger
values of $\ell $. The results obtained with the CMS2 limit are
uninteresting hence not reported: indeed, the values of $p$ which minimize
the value of the limit are either $p=1/2$ for small values of $\ell $ (in
which case this limit is analogous but less stringent than the CC limit, see
(\ref{CC})), or $p=1$ for larger values of $\ell $ (and this yields the BS$%
\ell $ limit). The new limits NUL2$\ell $ and NLL2$\ell $ clearly yield the
most stringent results. The NLL1n$\ell $ lower limit only yields cogent
results for large values of the angular momentum. The NLL3 lower limit works
reasonably well for small values of $\ell $ but becomes poor for higher
values of the angular momentum. The limits of the second kind ULSK and LLSK
yield similar results to those given by the NUL2$\ell $ and NLL2$\ell $
limits. Note that the arbitrary radii $r_{0}^{(\text{lo,incr)}}$ and $r_{0}^{(\text{lo,decr)}}$ have been chosen to optimize the restriction on the number of $\ell$-wave bound states. Finally, the results obtained with the C$\ell $n and the NLL4 lower
limits are not reported because they are very poor. These limits give $%
N_{\ell }\geq 1$ for small value of $\ell $ and $N_{\ell }\geq 0$ for large
value of $\ell $. This defect comes from the presence of the factor $%
1/(2\ell +1)$ which for instance implies that this lower bound becomes three
times smaller when $\ell $ go from $0$ to $1$ while the actual number of
bound states $N_{\ell }$ decreases generally only by one or two units.

The last test is performed with the Y potential (\ref{yuk}). The exact
number $N_{\ell }$ of bound states is again computed by integrating
numerically (\ref{EqEta1}) with $\eta (0)=0$ and $\eta (\infty )=N_{\ell
}\pi $, see Section \ref{sec3}.

The NUL1$\ell $ limit takes the form 
\begin{equation}
\text{NUL1}\ell :\quad N_{\ell }<1+\frac{2}{\pi }\sqrt{x_{+}-x_{-}}\left\{
g^{2}\int_{x_{-}}^{x_{+}}dx\frac{\exp (-x)}{x}-\ell (\ell +1)\left( \frac{1}{%
x_{-}}-\frac{1}{x_{+}}\right) \right\} ^{1/2},  \label{nul1_yuk}
\end{equation}%
where $x_{\pm }$ are the two solutions of the following equation 
\begin{equation}
\ell (\ell +1)=g^{2}\,x_{\pm }\,\exp (-x_{\pm }).  \label{xpm2}
\end{equation}

The NLL1n$\ell $ limit reads 
\begin{equation}
\text{NLL1n}\ell :\quad N_{\ell }>-1+\frac{1}{\pi }\left\{
g^{2}\int_{x_{-}}^{x_{+}}dx\frac{\exp (-x)}{x}-\ell (\ell +1)\left( \frac{1}{%
x_{-}}-\frac{1}{x_{+}}\right) \right\} \ |V_{\ell ,\text{eff}}^{\text{min}%
}|^{-1/2},  \label{nll1_yuk}
\end{equation}%
where $V_{\ell ,\text{eff}}^{\text{min}}$ is the minimum value of the
effective potential (\ref{Veff}). The NUL2$\ell $ and NLL2$\ell $ limits can
be written as follows: 
\begin{eqnarray}
\text{NUL2}\ell :\quad N &<&\frac{1}{\pi }G(g,\ell ;x_{-},x_{+})+\frac{1}{%
2\pi }\log \left\vert \frac{V_{\ell ,\text{eff}}^{\text{min}}}{M}\right\vert
+1,  \label{nul2_yuk} \\
\text{NLL2}\ell :\quad N &>&\frac{1}{\pi }G(g,\ell ;x_{-},x_{+})-\frac{1}{%
2\pi }\log \left\vert \frac{V_{\ell ,\text{eff}}^{\text{min}}}{M}\right\vert
-\frac{3}{2},
\end{eqnarray}%
where 
\begin{equation}
G(g,\ell ;a,b)=\int_{a}^{b}\frac{dx}{x}\sqrt{g^{2}x\exp (-x)-\ell (\ell +1)}.
\label{gab}
\end{equation}%
and 
\begin{subequations}
\begin{equation}
M=\min \left[ |V_{\ell ,\text{eff}}(p)|,|V_{\ell ,\text{eff}}(q)|\right] 
\end{equation}%
where $p$ and $q$ are solutions of 
\begin{equation}
G(g,\ell ;x_{-},\frac{p}{R})=\frac{\pi }{2},\quad G(g,\ell ;\frac{q}{R}%
,x_{+})=\frac{\pi }{2}.
\end{equation}

The lower limits NLL3 and NLL4 take somewhat simpler forms: 
\end{subequations}
\begin{equation}
\text{NLL3}:\quad N_{\ell }>\sqrt{\frac{2}{\pi }}\,g-\frac{x^{2}-y^{2}}{2\pi 
}\log \left( \frac{4g}{\pi }\right) -\frac{3}{2}-\frac{(1+4\ell )}{2\pi }%
\log \left( \frac{x}{y}\right) ,  \label{nul3_yuk}
\end{equation}%
where $y$ and $x$ are defined by $\text{erf}(y)=\sqrt{\pi /8}/g$, $\text{erf}%
(x)=1-\sqrt{\pi /8}/g$, and 
\begin{equation}
\text{NLL4}:\quad N_{\ell }>\frac{g}{\pi \sqrt{e}(2\ell +1)}-\frac{1}{2}.
\label{nul4_yuk}
\end{equation}

The previously known limits take the form: 
\begin{equation}
\text{BS}\ell :\quad N_{\ell }<\frac{g^{2}}{2\ell +1};
\end{equation}%
\begin{equation}
\text{CMS}:\quad N_{\ell }<2\sqrt{\frac{2}{\pi }}g+1-\sqrt{1+\frac{4\,\ell
\,(\ell +1)}{\pi ^{2}}};
\end{equation}%
\begin{subequations}
\begin{equation}
\text{M}\ell :\quad N_{\ell }<(AB)^{1/4}
\end{equation}%
where $A$ and $B$ are given by 
\begin{eqnarray}
A &=&g^{2}(\exp (-x_{-})(1+x_{-})-\exp (-x_{+})(1+x_{+}))-\ell (\ell
+1)(x_{+}-x_{-}), \\
B &=&g^{2}\int_{x_{-}}^{x_{+}}dx\frac{\exp (-x)}{x}-\ell (\ell +1)\left( 
\frac{1}{x_{-}}-\frac{1}{x_{+}}\right) ;
\end{eqnarray}%
\end{subequations}
\begin{equation}
\text{GGMT}:\quad N_{\ell }\leq g^{2p}(2\ell +1)^{(1-2p)}\,\frac{C_{p}\Gamma
(p)}{p^{p}};
\end{equation}%
\begin{equation}
\text{CMS2}:\quad N_{\ell }\leq g^{2p}(2\ell +1)^{(1-2p)}\,\frac{\tilde{C}%
_{p}\Gamma (p)}{p^{p}};
\end{equation}%
\begin{equation}
\text{C}\ell :\quad N_{\ell }\geq \frac{2g}{\pi (2\ell +1)}\,\sqrt{y}\exp \left(-\frac{y}{2}\right)-\frac{1}{2},
\end{equation}%
where $y$ is the solution of $\exp (-y)=(2\ell +1)y^{2\ell }\Gamma (-2\ell
,y)$.

\begin{table}[tbp]
\caption{Comparison for the Y potential (\protect\ref{yuk}) between the
exact number $N_{\ell}$ of bound states, various upper and lower limits on $%
N_{\ell}$ previously known and new upper and lower limits on $N_{\ell}$, for
several values of $g$ and $\ell$.}
\label{tab2}
\begin{center}
\begin{tabular}{cc|cccccccccccc}
$g$ & $\ell $ & LLSK & NLL3 & NLL1n$\ell$ & NLL2$\ell $ & Ex & NUL2$\ell $ & 
BS$\ell $ & CMS & M$\ell $ & GGMT & NUL1$\ell $ & ULSK \\ \hline
8 & 1 & 3 & 3 & 2 & 3 & \textbf{5} & 6 & 21 & 12 & 8 & 19 & 16 & 7 \\ 
& 3 & 1 & 0 & 2 & 1 & \textbf{2} & 3 & 9 & 11 & 5 & 4 & 5 & 5 \\ 
15 & 2 & 7 & 5 & 4 & 7 & \textbf{9} & 11 & 45 & 23 & 17 & 41 & 32 & 13 \\ 
& 6 & 2 & 0 & 3 & 2 & \textbf{4} & 5 & 17 & 20 & 8 & 8 & 9 & 8 \\ 
22 & 3 & 11 & 8 & 5 & 11 & \textbf{13} & 15 & 69 & 33 & 25 & 64 & 48 & 17 \\ 
& 9 & 4 & 0 & 4 & 4 & \textbf{5} & 6 & 25 & 29 & 12 & 12 & 13 & 11 \\ 
29 & 4 & 16 & 10 & 7 & 16 & \textbf{18} & 19 & 93 & 44 & 33 & 87 & 64 & 22
\\ 
& 12 & 5 & 0 & 5 & 6 & \textbf{7} & 8 & 33 & 39 & 16 & 16 & 18 & 14 \\ 
35 & 5 & 19 & 11 & 8 & 19 & \textbf{21} & 23 & 111 & 53 & 40 & 104 & 76 & 26
\\ 
& 15 & 6 & 0 & 6 & 6 & \textbf{8} & 9 & 39 & 46 & 18 & 18 & 20 & 16 \\ 
41 & 6 & 23 & 12 & 10 & 23 & \textbf{25} & 27 & 129 & 62 & 47 & 120 & 89 & 30
\\ 
& 18 & 7 & 0 & 6 & 7 & \textbf{9} & 10 & 45 & 54 & 20 & 20 & 22 & 18 \\ 
\hline
\end{tabular}%
\end{center}
\end{table}

Comparisons between the various limits and the exact results are presented
in Table \ref{tab2}. The characteristics of the various limits for the Y
potential are analogous to those commented above for the E potential. Here
again the NUL2$\ell $ and the NLL2$\ell $ limits are the most effective
ones, being indeed fairly stringent for all the values of $g$ and $\ell $
considered, and the limits of the second kind ULSK and LLSK also give quite
stringent limits.

\subsection{Tests of the limits for the value of $L$}

\label{sec2.2}

In this subsection we test various limits on the largest value $L$ of the
angular momentum quantum number $\ell $ for which the potentials E and Y do
possess bound states. In this article we only obtained \textit{new }lower
limits on $L$. Indeed neat upper limits on $L$ cannot be extracted from the 
\textit{new} upper limits on $N_{\ell }$ presented in Section \ref{sec1}.
But we will now see that the \textquotedblleft na\"{\i}ve\textquotedblright\
upper limit $L_{\text{eff}}^{(+)}$ (\ref{ULL}) is quite good indeed better
than the previously \textit{known} upper limits $L_{\text{BSL}}^{(+)}$ and $%
L_{\text{CMSL}}^{(+)},$ see (\ref{BS2}) and (\ref{CMSL}).

The first test is performed with the nonsingular E potential (\ref{expo}).
The lower limit NLL3L gives 
\begin{subequations}
\label{lexp3}
\begin{equation}
L_{\text{NLL3L}}^{(-)}=\left\{ \left\{ \frac{\nu }{\lambda }\right\}
\right\} ,  \label{lexp3a}
\end{equation}%
with 
\begin{eqnarray}
\nu  &=&\frac{2g}{\pi }-\frac{1}{2\pi }\log \left[ \frac{1-x}{x}\right] -%
\frac{3}{2},  \label{nuexp3} \\
\lambda  &=&\frac{1}{\pi }\log \left[ \frac{\log x}{\log (1-x)}\right] ,
\end{eqnarray}%
and $x=\pi /(4g)$. The lower limit NLL4L gives 
\end{subequations}
\begin{equation}
L_{\text{NLL4L}}^{(-)}=\left\{ \left\{ \frac{1}{2}\left( \frac{4g}{\pi e}%
-1\right) \right\} \right\} .  \label{lexp4}
\end{equation}

The previously known upper limits on $L$ can also be obtained analytically: 
\begin{equation}  \label{BSLexp}
L_{\text{BSL}}^{(+)}=\left\{ \left\{ \frac{1}{2}\left(g^{2}-1\right)\right\}
\right\},
\end{equation}
\begin{equation}  \label{CMSLexp}
L_{\text{CMSL}}^{(+)}=\left\{ \left\{ \frac{1}{2}\left(4\,g-1\right)\right\}
\right\},
\end{equation}
\begin{equation}  \label{ULLexp}
L_{\text{eff}}^{(+)}=\left\{ \left\{ \frac{1}{2}\left(\frac{4g}{e}%
-1\right)\right\} \right\}.
\end{equation}

\begin{figure}
\begin{center}
\includegraphics[width=10cm]{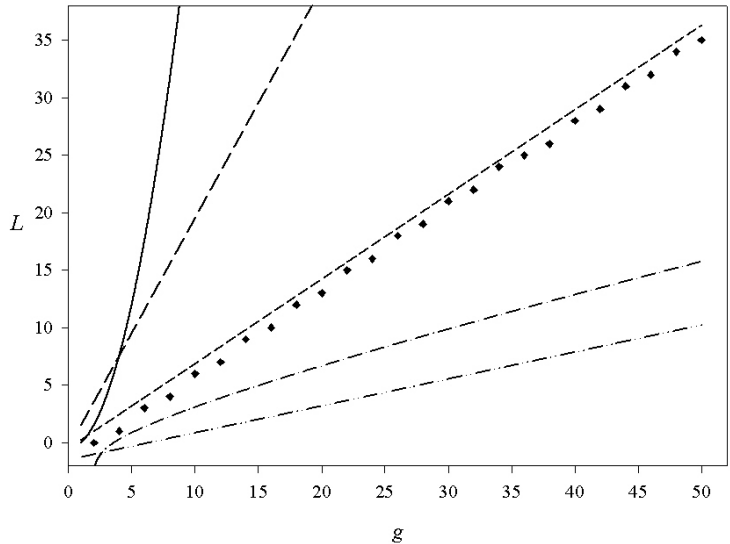}
\end{center}
\caption{Comparison between the exact value of $L$ (diamond), the upper
limits BSL (\protect\ref{BSLexp}) (solid), CMSL (\protect\ref{CMSLexp})
(long dash) and $L_{\text{eff}}^{(+)}$ (\protect\ref{ULLexp}) (short dash)
and the lower limits NLL3L (\protect\ref{lexp3}) (dash-dot) and NLL4L (%
\protect\ref{lexp4}) (dash-dot-dot) for the E potential (\protect\ref{expo}).
}
\label{fig2}
\end{figure}

A comparison between these limits and the exact results (computed
numerically, as indicated in the preceding subsection) is presented in Fig. %
\ref{fig2}. Except for the BSL upper limit, we have the correct linear
behavior in $g$ as discussed above, the best result being clearly provided
by the na\"{\i}ve upper bound $L_{\text{eff}}^{(+)}$. Indeed this result
appears hardly improvable, because the error introduced by this upper limit $%
L_{\text{eff}}^{(+)}$ is of at most one unit (at least for this example, as
well as the following one, see below).

The second test is performed with the singular Y potential (again the exact
result can only be computed numerically). The lower limit NLL3L gives 
\begin{subequations}
\label{lyuk3}
\begin{equation}
L_{\text{NLL3L}}^{(-)}=\left\{ \left\{ \frac{\nu }{\lambda }\right\}
\right\} ,
\end{equation}%
with 
\begin{eqnarray}
\nu &=&\sqrt{\frac{2}{\pi }}g-\frac{x^{2}-y^{2}}{2\pi }-\frac{1}{2\pi }\log %
\left[ \frac{x}{y}\right] -\frac{3}{2},  \label{nuyuk3} \\
\lambda &=&\frac{2}{\pi }\log \left[ \frac{x}{y}\right] ,
\end{eqnarray}%
where $y$ and $x$ are defined by $\text{erf}(y)=\alpha $, $\text{erf}%
(x)=1-\alpha $, $\alpha =\sqrt{(\pi /8)}/g$. The lower limit NLL4L gives 
\end{subequations}
\begin{equation}
L_{\text{NLL4L}}^{(-)}=\left\{ \left\{ -\frac{1}{2}+\frac{g}{\pi \sqrt{e}}%
\right\} \right\} .  \label{lyuk4}
\end{equation}%
The previously known limits on $L$ can also be obtained analytically: 
\begin{equation}  \label{BSLyuk}
L_{\text{BSL}}^{(+)}=\left\{ \left\{ \frac{1}{2}\left(g^{2}-1\right)\right\}
\right\},
\end{equation}
\begin{equation}  \label{CMSLyuk}
L_{\text{CMSL}}^{(+)}=\left\{ \left\{ \frac{1}{2}\left(4\sqrt{\frac{2}{\pi }}%
\,g-1\right)\right\} \right\},
\end{equation}
\begin{equation}  \label{ULLyuk}
L_{\text{eff}}^{(+)}=\left\{ \left\{ \frac{1}{2}\left(\frac{2g}{\sqrt{e}}%
-1\right)\right\} \right\}.
\end{equation}

\begin{figure}
\begin{center}
\includegraphics[width=10cm]{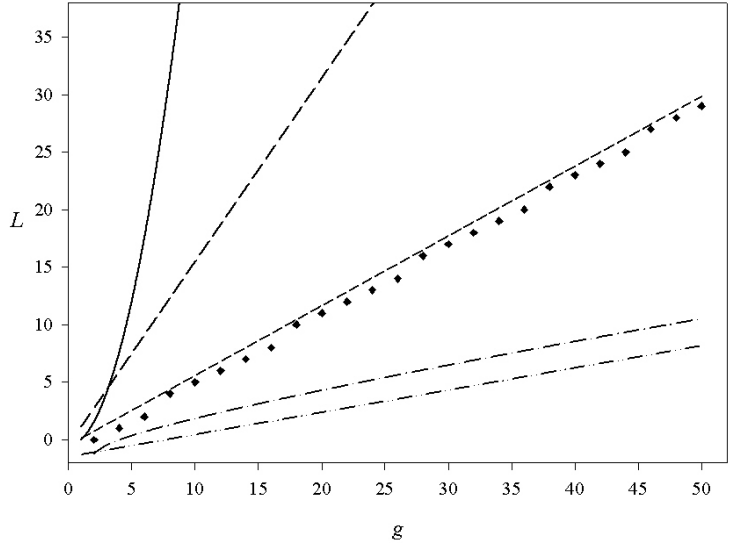}
\end{center}
\caption{Comparison between the exact value of $L$ (diamond), the upper
limits BSL (\protect\ref{BSLyuk}) (solid), CMSL (\protect\ref{CMSLyuk})
(long dash) and $L_{\text{eff}}^{(+)}$ ( \protect\ref{ULLyuk}) (short dash)
and the lower limits NLL3L (\protect\ref{lyuk3}) (dash-dot) and NLL4L (%
\protect\ref{lyuk4}) (dash-dot-dot) for the Y potential (\protect\ref{yuk}).}
\label{fig3}
\end{figure}

The comparison between these limits and the exact results is presented in
Fig. \ref{fig3}. Here again, the best result is obtained from the na\"{\i}ve
upper bound $L_{\text{eff}}^{(+)}$ for which the error is, as indicated
above, of at most one unit.

\subsection{Tests of the limits for the total number of bound states $N$}

\label{sec2.3}

This subsection is devoted to test the limits on the total number of bound
states $N$. We will not test the BiS limit due to its bad behavior at large $%
g$. We do however test the BSN limit which yields the same incorrect
behavior but is simpler to compute.

The first test is performed with the E potential (\ref{expo}). Again, the
exact result can only be calculated numerically.

The \textit{new }upper limit NUL2Nm, see (\ref{NUL2Nb}), takes the simple
form 
\begin{equation}
N<\frac{1}{8}\left( \frac{4g}{e}+1\right) ^{2}\,\left( \frac{4g}{\pi }+\frac{%
1}{\pi }\log \frac{4g}{\pi }+1\right) .  \label{NUL2Nmexp}
\end{equation}

The \textit{new} lower limit NLLN3 reads 
\begin{equation}
N>\frac{\nu }{6}\left( 2\left( L_{\text{NLL3L}}^{(-)}\right) ^{2}+7L_{\text{%
NLL3L}}^{(-)}+6\right) ,  \label{nlln3exp}
\end{equation}%
with $L_{\text{NLL3L}}^{(-)}$ and $\nu $ given by equations (\ref{lexp3a})
and (\ref{nuexp3}); and the \textit{new }lower limit NLLN4 is given by (\ref%
{NLL3N}) with $\sigma =4\,g/(e\,\pi )$.

The other, previously known limits read as follows: 
\begin{equation}  \label{BSNexp}
\text{BSN}:\quad N<\frac{1}{2}g^{2}(g^{2}+1),
\end{equation}
\begin{equation}  \label{Lexp}
\text{Lcentral}:\quad N<0.864\,g^{3},
\end{equation}
\begin{equation}  \label{CMSNexp}
\text{CMSN}:\quad N<1.698\,(g^{3}+2.3562\,g^{2}+1.116\,g+0.1473).
\end{equation}

\begin{figure}
\begin{center}
\includegraphics[width=10cm]{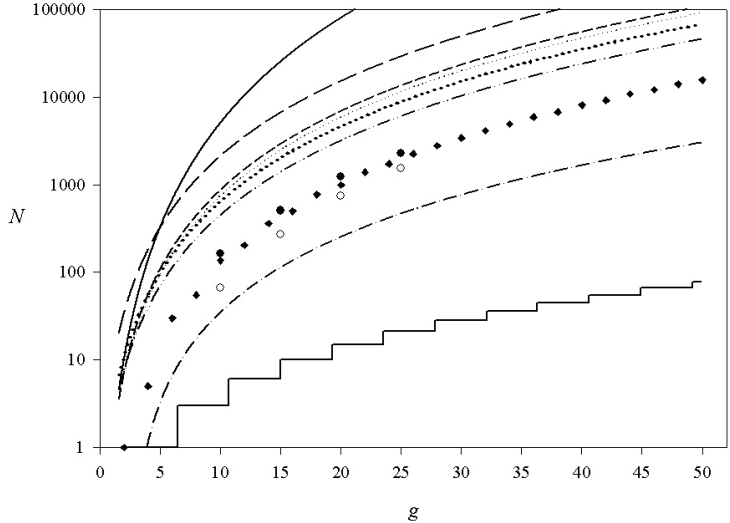}
\end{center}
\caption{Comparison between the exact value of $N$ (diamond), the upper
limits BSN (\protect\ref{BSNexp}) (solid), CMSN (\protect\ref{CMSNexp})
(long dash), Lcentral (\protect\ref{Lexp}) (short dash), \textit{improved}
BSN (\protect\ref{BSNexpimp}) (dot), \textit{improved} CMSN (\protect\ref{CMSNexpimp}) (sparse dot), NUL2Nm (\protect\ref%
{NUL2Nmexp}) (dash dot dot) and the lower limits NLLN3 (\protect\ref%
{nlln3exp}) (dash dot) and NLLN4 (\protect\ref{NLL3N}) (ladder) for the E
potential (\protect\ref{expo}). The black and the white circles correspond
respectively to the NUL2N and NLL2N limits.}
\label{fig4}
\end{figure}

The BSN and the CMSN limits can be improved: instead of using the limit on $L
$ provided by these limits ($L_{\text{BSL}}^{(+)}$ and $L_{\text{CMSL}}^{(+)}
$), we can use the best upper limits $L_{\text{eff}}^{(+)}$ (\ref{ULL}); the BSN and
CMSN limits obtained in this manner are called here \textit{improved} BSN
and CMSN limits:
\begin{equation}
\label{BSNexpimp}
\text{\textit{improved} BSN:}\quad N<\frac{1}{2}g^2 \left(\frac{4g}{e}+1\right)
\end{equation}
\begin{equation}
\label{CMSNexpimp}
\text{\textit{improved} CMSN:}\quad N<0.5202\,(g^{3}+2.179\,g^{2}+1.726\,g+0.4806)
\end{equation}

Fig. \ref{fig4} presents a comparison between the various limits and the
exact result. It shows that the limits on the \textit{total} number of bound
states which can be expressed in a neat form are not very stringent.
[Indeed, the best results are yielded by the upper limit NUL2Nn which is
obtained using only a limit on the number of S-wave bound states, $N_{0}$,
and the simple limit $L_{\text{eff}}^{(+)}$ on the maximal value $L$ of $%
\ell $ for which bound states do exist]. There are at least three reasons
for this. First, most of the limits do not contain the appropriate
functional of the potential (as identified by the asymptotic behavior at
large $g$ of $N,$ see (\ref{NpropG3b})): only the Lieb limit Lcentral, see (%
\ref{Lc}), features the correct form, but the numerical factor is not
optimal indeed too large (by approximately a factor 7). The second reason is
that for every value of $\ell $, there is a round-off error introduced by
the limit; to obtain the limit on the total number of bound states we sum
all these errors. The third reason is that to be able to make the summation
over the values of $\ell $ we must have an explicit dependence of the limits
on $\ell $, and this entails that we cannot use some of the limits we found;
in particular we cannot use the \textit{new }upper and lower limits NUL2 and
the NLL2, which are quite stringent, to obtain a neat formula. But we can
use them and compute upper and lower limits on $N_{\ell }$, then sum all
these contributions to obtain upper and lower limits on $N$, the sum being
stopped when $N_{\ell }^{(+)}$ is smaller than 1 and $N_{\ell }^{(-)}$ is
negative (see subsection \ref{sec1.6}). We call NUL2N respectively NLL2N the
upper respectively lower limits on the total number of bound states $N$\
obtained (from NUL2 respectively NLL2) via this (inelegant) procedure. Fig. %
\ref{fig4} shows, for 4 values of $g$, that these limits are quite stringent.

The second test is performed with the Y potential. Again, the exact total
number of bound states is computed numerically, as indicated above.

The \textit{new }upper limit NUL2Nm, see (\ref{NUL2Nb}) reads%
\begin{equation}
\text{NUL2Nm:}\quad N<\frac{1}{8}\left( \frac{2g}{\sqrt{e}}+1\right)
^{2}\,\left( 2g\sqrt{\frac{2}{\pi }}+\frac{x^{2}-y^{2}}{\pi }+\frac{1}{\pi }%
\log \frac{x}{y}+1\right) ,  \label{NUL2Nmyuk}
\end{equation}%
where $y$ and $x$ are defined by $\text{erf}(y)=\sqrt{\pi /8}/g$, $\text{erf}%
(x)=1-\sqrt{\pi /8}/g$. The \textit{new }limits NLLN3 and NLLN4 are: 
\begin{equation}
\text{NLLN3:}\quad N>\frac{\nu }{6}\left( 2\left( L_{\text{NLL3L}%
}^{(-)}\right) ^{2}+7L_{\text{NLL3L}}^{(-)}+6\right) ,  \label{nlln3yuk}
\end{equation}%
with $L_{\text{NLL3L}}^{(-)}$ and $\nu $ given by equations (\ref{lyuk3})
and (\ref{nuyuk3}); while the lower limit NLLN4 is given by (\ref{NLL3N})
with $\sigma =2\,g/(\sqrt{e}\pi)$.

The other previously known limits read: 
\begin{equation}  \label{BSNyuk}
\text{BSN}:\quad N<\frac{1}{2}g^{2}(g^{2}+1),
\end{equation}
\begin{equation}  \label{Lyuk}
\text{Lcentral}:\quad N<0.703\,g^{3},
\end{equation}
\begin{equation}  \label{CMSNyuk}
\text{CMSN}:\quad N<3.3422\,(g^{3}+1.88\,g^{2}+0.711\,g+0.075).
\end{equation}
\begin{equation}
\label{BSNyukimp}
\text{\textit{improved} BSN:}\quad N<\frac{1}{2}g^2 \left( \frac{2g}{\sqrt{e}}+1\right)
\end{equation}
\begin{equation}
\label{CMSNyukimp}
\text{\textit{improved} CMSN:}\quad N<0.4924\,(g^{3}+2.237\,g^{2}+1.781\,g+0.5078)
\end{equation}

\begin{figure}
\begin{center}
\includegraphics[width=10cm]{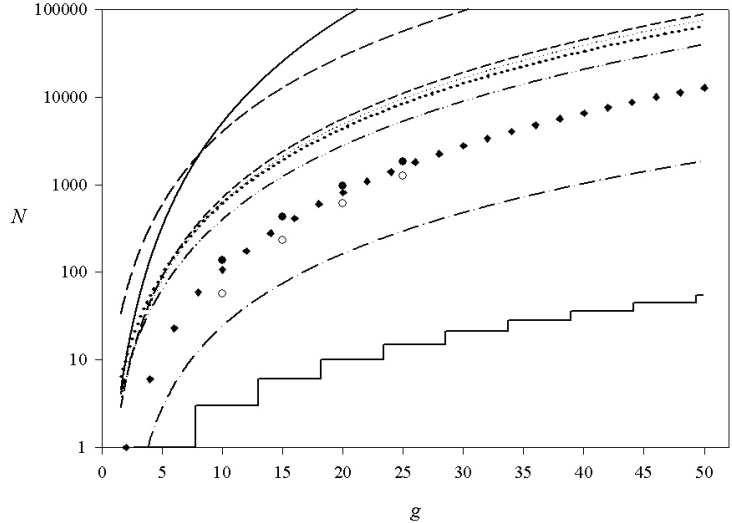}
\end{center}
\caption{Comparison between the exact value of $N$ (diamond), the upper
limits BSN (\protect\ref{BSNyuk}) (solid), CMSN (\protect\ref{CMSNyuk})
(long dash), Lcentral (\protect\ref{Lyuk}) (short dash), \textit{improved}
BSN (\protect\ref{BSNyukimp}) (dot) and \textit{improved} CMSN (\protect\ref{CMSNyukimp}) (sparse dot), NUL2Nm (\protect\ref{NUL2Nmyuk}) (dash dot dot) and the lower limits NLLN3 (\protect\ref{nlln3yuk}) (dash dot) and NLLN4 (\protect\ref{NLL3N}) (ladder) for the Y potential (\protect\ref{yuk}). The black and the white circles correspond
respectively to the NUL2N and NLL2N limits.}
\label{fig5}
\end{figure}

A comparison between the various limits and the exact numerical results is
presented in Fig. \ref{fig5}, in analogy to the case of the E potential and
with analogous conclusions, see above.

\section{PROOFS}

\label{sec3}

In this section we prove the \textit{new} results reported in Section \ref%
{sec1}. Because we tried and presented those results in Section \ref{sec1}
in a user-friendly order, the proofs given below do not follow the same
order, due to the need here to follow a more logical sequence. To provide
some guidance we divided this section into several subsections, but we must
forewarn the reader that a sequential reading is essential to understand
what goes on.

\subsection{Tools}

\label{sec3.1}

The starting point of our treatment is the well-known fact (see Section \ref%
{sec1}) that the number $N_{\ell }$ of ($\ell $-wave) bound states possessed
by the potential $V(r)$ coincides with the number of zeros, in the interval $%
0<r<\infty ,$ of the function $u(r)$\ uniquely defined (up to an irrelevant
multiplicative constant) as the solution of the zero-energy $\ell $-wave
radial Schr\"{o}dinger equation 
\begin{subequations}
\label{ZESch}
\begin{equation}
u^{\prime \prime }(r)=\left[ V(r)+\frac{\ell (\ell +1)}{r^{2}}\right] \,u(r)
\label{ZEScha}
\end{equation}%
with the boundary condition 
\begin{equation}
u(0)=0.  \label{B0}
\end{equation}

To get an efficient handle on the task of counting these zeros (or rather,
of providing upper and lower limits on their number) it is convenient to
introduce a new dependent variable $\eta (r)$ related to $u(r)$\ as follows: 
\end{subequations}
\begin{equation}
\left[ -U(r)\right] ^{1/2}\cot \left[ \eta (r)\right] =f(r)+\frac{u^{\prime
}(r)}{u(r)}.  \label{DefEta}
\end{equation}%
Here we moreover introduce two new functions: a function $f(r)$, which we
reserve to assign at our convenience below with the only proviso that it be
finite in the open interval $0<r<\infty $, and nonnegative throughout this
interval, 
\begin{equation}
f(r)\geq 0\quad \text{for}\quad 0\leq r<\infty ;
\end{equation}%
and a function $U(r),$ which might or might not coincide with the potential $%
V(r)$ (see below) but that (unless we explicitly state otherwise) we require
to be finite in the open interval $0<r<\infty $ , to satisfy (at least) the
properties (see (\ref{V2}), (\ref{VatInf}) and (\ref{VatZero})) 
\begin{equation}
-U(r)=\left\vert U(r)\right\vert ,  \label{Unonneg}
\end{equation}%
\begin{equation}
\underset{r\rightarrow \infty }{\lim }\left[ r^{2+\varepsilon }U(r)\right]
=0,  \label{UatInf}
\end{equation}%
\begin{equation}
\underset{r\rightarrow 0}{\lim }\left[ r^{2-\varepsilon }U(r)\right] =0,
\label{UatZero}
\end{equation}%
and to be related to the potential $V(r)$ as follows: 
\begin{equation}
U(r)+W(r)=V(r)+\frac{\ell (\ell +1)}{r^{2}},  \label{DefVmin}
\end{equation}%
where we still reserve the privilege to assign the function $W(r)$ at our
convenience (a possibility will be to set $U(r)=V(r)$ hence $W(r)=\ell (\ell
+1)/r^{2}$;$\,$but it shall not be the only one, see below). Of course the
function $\eta (r)$ (as well as $u(r)$) depends on $\ell $, although for
notational simplicity we omit to indicate this explicitly, and this remark
may as well apply to the other functions, $f(r),$ $U(r),$ $W(r)$, introduced
here and utilized below.

It is then easy to see that the function $\eta (r)$ is uniquely
characterized by the first-order nonlinear ODE (implied by (\ref{DefEta})
with (\ref{DefVmin}), and (\ref{ZEScha})) 
\begin{eqnarray}
\eta ^{\prime }(r) &=&\left\vert U(r)\right\vert ^{1/2}+\left\vert
U(r)\right\vert ^{-1/2}\left\{ \left[ f(r)\right] ^{2}-f^{\prime}(r)-W(r)%
\right\} \sin ^{2}\left[ \eta (r)\right]  \notag \\
&&-\left\{ \left[ 4\left\vert U(r)\right\vert \right] ^{-1}\,U^{%
\prime}(r)+f(r)\right\} \sin \left[ 2\eta (r)\right] ,  \label{EqEta}
\end{eqnarray}
with the boundary condition (implied by (\ref{DefEta}) with (\ref{B0}) and (%
\ref{UatZero})) 
\begin{equation}
\eta (0)=0.  \label{EtaZero}
\end{equation}
We of course assume the function $\eta (r)$ to be continuous, disposing
thereby of the $\text{mod}(\pi )$ ambiguity entailed by the definition (\ref%
{DefEta}).

It is now easy to see that this function $\eta (r)$ provides a convenient
tool to evaluate the number of zeros $N_{\ell }$ of $u(r)$. Indeed, if we
denote with $z_{n}$ the zeros of $u(r),$ $u(z_{n})=0$, ordered so that 
\begin{equation}
0\equiv z_{0}<z_{1}<\ldots <z_{N_{\ell }}
\end{equation}%
(see Section \ref{sec1}), since clearly (see (\ref{EqEta})) whenever $\eta
(r)$ is an integer multiple of $\pi $ the derivative $\eta ^{\prime }(r)$ is
nonnegative, $\eta ^{\prime }(r)\geq 0$, we may conclude (see (\ref{DefEta}%
)) that 
\begin{equation}
\eta (z_{n})=n\pi ,\quad n=0,1,\ldots ,N_{\ell },  \label{EtaAtEvenPi}
\end{equation}%
with 
\begin{equation}
(n-1)\pi \leq \eta (r)\leq n\pi \quad \text{for}\quad z_{n-1}\leq r\leq
z_{n},\ n=1,\ldots ,N_{\ell }.  \label{Eta1}
\end{equation}%
It is moreover plain that, provided (see (\ref{DefEta})) 
\begin{equation}
\underset{r\rightarrow \infty }{\lim }\left[ \frac{\left\vert
U(r)\right\vert ^{1/2}}{f(r)+\frac{u^{\prime }(r)}{u(r)}}\right] =0,
\label{condU}
\end{equation}%
there holds the asymptotic relation 
\begin{equation}
\eta (\infty )=N_{\ell }\pi ,  \label{EtaInf}
\end{equation}%
and that this asymptotic value is approached from above. [To prove the last
statement one sets, in the asymptotic large $r$ region, $\eta (r)=N_{\ell
}\pi +\varepsilon (r)$ with $\left\vert \varepsilon (r)\right\vert \ll 1$,
and uses the asymptotic estimates $\sin ^{2}(N_{\ell }\pi +\varepsilon
)\approx \varepsilon ^{2},$ $\sin \left[ 2(N_{\ell }\pi +\varepsilon )\right]
\approx 2\varepsilon $ to rewrite the ODE (\ref{EqEta}) in the asymptotic
region as follows (recall that $U(r)$ is nonpositive, see (\ref{Unonneg})): 
\begin{equation}
\varepsilon ^{\prime }(r)\approx \left[ -U(r)\right] ^{1/2}+\left\{ \frac{%
U^{\prime }(r)}{2\,U(r)}-2\,f(r)\right\} \varepsilon (r).
\end{equation}%
One can then replace the approximate equality sign $\approx $ in this
formula with the equality sign = and integrate the resulting linear ODE,
obtaining 
\begin{equation}
\varepsilon (r)=\left\vert U(r)\right\vert ^{1/2}\int_{R}^{r}dx\,\exp \left[
2\int_{r}^{x}dy\,f(y)\right] ,
\end{equation}%
where $R$ is a (finite) integration constant, and this formula (valid in the
asymptotic, large $r$, region, where of course $r>R$) shows that $%
\varepsilon (r)$ is indeed positive].

To get a more detailed information on the behavior of the function $\eta (r)$
we make the additional assumption that the assignments of the auxiliary
functions $f(r)$ and $W(r)$ guarantee validity of the following inequality: 
\begin{equation}
\left[ f(r)\right] ^{2}-f^{\prime }(r)-W(r)\geq 0  \label{fW}
\end{equation}%
(it would be enough for our purposes that this inequality be valid only at
large values of $r$ -- but for simplicity we assume hereafter its validity
for all values of $r,$ $0\leq r<\infty $). It is then clear from the ODE (%
\ref{EqEta}) satisfied\ by $\eta (r)$ that, wherever $\eta (r)$ takes a
value which is an odd integer multiple of $\pi /2$, namely at the points $%
r=b_{n}$ such that 
\begin{equation}
\eta (b_{n})=\frac{1}{2}(2n-1)\pi ,\quad n=1,2,\ldots ,N_{\ell },
\label{EtaAtOddPi}
\end{equation}%
its derivative $\eta ^{\prime }(r)$ is nonnegative, $\eta ^{\prime
}(b_{n})\geq 0$. Hence we may complement the information provided by (\ref%
{EtaAtEvenPi}) with that provided by this formula, (\ref{EtaAtOddPi}), and
moreover replace the information provided by (\ref{Eta1}) with the following
more detailed information: 
\begin{equation}
(n-1)\pi \leq \eta (r)\leq \frac{1}{2}(2n-1)\pi \quad \text{for}\quad
z_{n-1}\leq r\leq b_{n},\ n=1,\ldots ,N_{\ell },  \label{BetaZ}
\end{equation}%
\begin{equation}
(2n-1)\frac{\pi }{2}\leq \eta (r)\leq n\pi \quad \text{for}\quad b_{n}\leq
r\leq z_{n},\ n=1,\ldots ,N_{\ell },  \label{ZetaB}
\end{equation}%
which of course also entails that the points $z_{n}$ and $b_{n}$ are
interlaced, 
\begin{equation}
z_{0}<b_{1}<z_{1}<b_{2}<\ldots <z_{N_{\ell }-1}<b_{N_{\ell }}<z_{N_{\ell
}}<\infty .  \label{OrderBZ}
\end{equation}%
And note in particular that these formulas entail the following important
inequality (implied by the non existence of $b_{N_{\ell }+1}$), valid for
all values of $r,$ $0\leq r<\infty $: 
\begin{equation}
0\leq \eta (r)<\left( N_{\ell }+\frac{1}{2}\right) \pi .  \label{Etauplo}
\end{equation}%
It is moreover clear that the maximum value of $\eta (r)$, 
\begin{equation}
\hat{\eta}=\underset{0\leq r<\infty }{\max }\left[ \eta (r)\right] ,
\label{EtaMax}
\end{equation}%
is actually attained in the interval $z_{N_{\ell }}<r<\infty $, and that it
lies in the range 
\begin{equation}
N_{\ell }\,\pi \leq \hat{\eta}<\left( N_{\ell }+\frac{1}{2}\right) \,\pi .
\label{IneqEta}
\end{equation}

We are now in the position to derive the \textit{new} upper and lower limits
reported in Section \ref{sec1}.

\subsection{Proof of the lower limits NLL4}

\label{sec3.2}

We begin by proving the lower limit NLL4, see (\ref{NLL3}). To this end we
assign as follows the functions $f(r)$ and $W(r)$: 
\begin{equation}
f(r)=\frac{\ell }{r},\quad W(r)=\frac{\ell (\ell +1)}{r^{2}},
\end{equation}
entailing that the left-hand side of the inequality (\ref{fW}) vanishes,
that 
\begin{equation}
U(r)=V(r),  \label{UeqV}
\end{equation}
that the definition (\ref{DefEta}) now reads 
\begin{equation}
\left\vert V(r)\right\vert ^{1/2}\,\cot \left[ \eta (r)\right] =\frac{\ell }{%
r}+\frac{u^{\prime }(r)}{u(r)},  \label{DefEta2}
\end{equation}
and, most importantly, that (\ref{EqEta}) reads 
\begin{equation}
\eta ^{\prime }(r)=\left\vert V(r)\right\vert ^{1/2}-\left\{ \frac{%
V^{\prime}(r)}{4\left\vert V(r)\right\vert }\,+\frac{\ell }{r}\right\} \sin %
\left[2\eta (r)\right] .  \label{EqEta1}
\end{equation}
Here we are of course assuming the potential $V(r)$ to satisfy the condition
(\ref{V2}).

Before proceeding with the proof, let us note that, for the potential (\ref%
{VeqR4}), the second term in the right-hand side of this ODE, (\ref{EqEta1}%
), vanishes, hence one immediately obtains 
\begin{equation}
\eta (r)=\frac{g}{2\ell +1}\min \left[ \alpha ^{2\ell +1},\left( \frac{r}{R}%
\right) ^{2\ell +1}\right] .
\end{equation}%
Hence (see (\ref{EtaInf})), for the potential (\ref{VeqR4}), 
\begin{equation}
N_{\ell }=\left\{ \left\{ \frac{g\,\alpha ^{2\ell +1}}{\pi (2\ell +1)}%
\right\} \right\} ,
\end{equation}%
where as usual the double braces signify that the integral part must be
taken of their contents. This observation implies that the upper and lower
limits which obtain (see below) by massaging the last term in the right-hand
side of the ODE (\ref{EqEta1}) are generally best possible, being saturated
by the potential (\ref{VeqR4}) (if need be, with an appropriate choice of
the parameter $\alpha $, see Section \ref{sec1}).

To prove (\ref{NLL3}) with (\ref{sigma}) we now introduce the auxiliary
function $\eta _{\text{lo}}(r)$ via the ODE

\begin{equation}
\eta _{\text{lo}}^{\prime }(r)=\left\vert V(r)\right\vert ^{1/2}-\left\{%
\frac{V^{\prime }(r)}{2\left\vert V(r)\right\vert }\,+\frac{2\ell }{r}%
\right\} \eta _{\text{lo}}(r)  \label{EqEtaLo}
\end{equation}
with the boundary condition 
\begin{equation}
\eta _{\text{lo}}(0)=0.  \label{EtaLoZero}
\end{equation}

We then assume the potential $V(r)$ to satisfy, in addition to (\ref{V2}),
the condition (\ref{CondV1}). It is then plain (see (\ref{EqEta1}) with (\ref%
{EtaZero}), (\ref{EqEtaLo}) with (\ref{EtaLoZero}), and (\ref{CondV1}))
that, for all values of $r,$ $0\leq r<\infty$, 
\begin{equation}
\eta _{\text{lo}}(r)\leq \eta (r).  \label{in1}
\end{equation}
[Indeed (\ref{EqEtaLo}) obtains from (\ref{EqEta1}) via the replacement $%
\sin (x)\Rightarrow x,$ and for positive $x$, $\sin (x)\leq x$; hence $\eta_{%
\text{lo}}(r)$ can never overtake $\eta (r)$ because, at the overtaking
point, a comparison of (\ref{EqEtaLo}) with (\ref{EqEta1}) entails $\eta _{%
\text{lo}}^{\prime}(r)\leq \eta^{\prime}(r)$, which negates the possibility
to perform the overtaking]. But the linear ODE (\ref{EqEtaLo}) with (\ref%
{EtaLoZero}) can be easily integrated to yield (recalling (\ref{V2})) 
\begin{equation}
\eta _{\text{lo}}(r)=\frac{r\,\left\vert V(r)\right\vert ^{1/2}}{2\ell +1},
\label{EtaLo}
\end{equation}
hence we conclude (see (\ref{EtaMax}), (\ref{in1}), (\ref{EtaLo}) and (\ref%
{sigma})) that 
\begin{equation}
\hat{\eta}\geq \frac{\pi \sigma }{2(2\ell +1)},
\end{equation}
and via (\ref{IneqEta}) this entails the lower limit NLL4, see (\ref{NLL3}),
which is thereby proven.

\subsection{Proof of the upper and lower limits NUL2 and NLL2}

\label{sec3.3}

Let us now prove the upper and lower limits NUL2 and NLL2, see (\ref{NUL2bis}%
) and (\ref{NLL2bis}), on the number $N_{0}$ of S-wave bound states
possessed by the central potential $V(r)$, of course under the assumption
that this potential satisfy the conditions (\ref{V+-}). The main tool of the
proof is the same function $\eta (r)$ as defined in the preceding subsection %
\ref{sec3.2}, which is therefore now defined by the formula (see (\ref%
{DefEta2})) 
\begin{equation}
\left\vert V(r)\right\vert ^{1/2}\,\cot \left[ \eta (r)\right] =\frac{%
u^{\prime }(r)}{u(r)},  \label{DefEta3}
\end{equation}
and satisfies the ODE (see (\ref{EqEta1})) 
\begin{equation}
\eta ^{\prime }(r)=\left\vert V(r)\right\vert ^{1/2}-\frac{V^{\prime }(r)}{%
4\left\vert V(r)\right\vert }\,\sin \left[ 2\eta (r)\right] .
\label{EqEtaZero}
\end{equation}
Note that this is just the function $\eta (r)$ that provided our main
analytical tool in \cite{BC}; however the conditions (\ref{V+-a}) and (\ref%
{V+-b}) (which clearly imply $V(r_{-})=0$) entail now, via (\ref{DefEta3}),
the condition 
\begin{equation}
\eta (r_{-})=0,
\end{equation}
as well as the fact that $u(r)$ is concave in the interval $0\leq r\leq
r_{-} $ (see (\ref{ZEScha}) and (\ref{V+-a})), hence it has no zero in that
interval, hence 
\begin{equation}
r_{-}<b_{1}<z_{1}  \label{LoLimToZ1}
\end{equation}
(see (\ref{EtaAtEvenPi}), (\ref{EtaAtOddPi}) and (\ref{DefEta3})). Likewise
the fact that $u(r)$ is also concave in the interval $r_{+}\leq r<\infty $
(see (\ref{ZEScha}) and (\ref{V+-d})), hence it has no extremum in that
interval, entails 
\begin{equation}
z_{N_{0}-1}<b_{N_{0}}<r_{+}.  \label{UpLimToZ(N-1)}
\end{equation}

To obtain the upper limit NUL2 we now integrate the ODE (\ref{EqEtaZero})
from $z_{1}$ to $z_{N_{0}-1}$ (and note that, thanks to (\ref{LoLimToZ1})
and (\ref{UpLimToZ(N-1)}), as well as (\ref{V+-}), we can hereafter replace,
whenever convenient, $\left\vert V(r)\right\vert $ with $-V^{(-)}(r)$, see (%
\ref{Vminus})): 
\begin{eqnarray}
\eta (z_{N_{0}-1})-\eta (z_{1})
&=&(N_{0}-2)\,\pi=\int_{z_{1}}^{z_{N_{0}-1}}dr\,\left[ -V^{(-)}(r)\right]
^{1/2}-\frac{1}{4} \int_{z_{1}}^{r_{\text{min}}}dr\,\frac{V^{\prime }(r)}{%
\left\vert V(r)\right\vert }\,\sin \left[ 2\eta (r)\right]  \notag \\
&&-\frac{1}{4}\int_{r_{\text{min}}}^{z_{N_{0}-1}}dr\,\frac{V^{\prime }(r)}{%
\left\vert V(r)\right\vert }\,\sin \left[ 2\eta (r)\right] .  \label{EtaPQ}
\end{eqnarray}
The first equality is of course entailed by (\ref{EtaAtEvenPi}). As for the
second equation, note that we conveniently split the integration of the
second term in the right-hand side of (\ref{EqEtaZero}) in two parts. The
properties (\ref{V+-b}), (\ref{V+-c}) (as well as the obvious fact that $%
\left\vert \sin (x)\right\vert \leq 1$), allow us to majorize the right-hand
side of this equation, (\ref{EtaPQ}). We thereby get 
\begin{equation}
\left( N_{0}-2\right) \,\pi <\int_{z_{1}}^{z_{N_{0}-1}}dr\,\left[ -V^{(-)}(r)%
\right] ^{1/2}+\frac{1}{4}\int_{z_{1}}^{r_{\text{min}}}dr\,\frac{%
V^{\prime}(r)}{V(r)}-\frac{1}{4}\int_{r_{\text{min}}}^{z_{N_{0}-1}}dr\,\frac{%
V^{\prime }(r)}{V(r)},
\end{equation}
hence, 
\begin{equation}
\left( N_{0}-2\right) \,\pi <\int_{z_{1}}^{z_{N_{0}-1}}dr\,\left[ -V^{(-)}(r)%
\right] ^{1/2}+\frac{1}{4}\,\log \left\{ \frac{\left[ V^{(-)}(r_{\text{min}})%
\right] ^{2}}{V^{(-)}(z_{1})\,V^{(-)}(z_{N_{0}-1})}\right\} .
\end{equation}
We now need to find quantities $p$ and $q$, defined only in terms of the
potential, such that $p\leq z_{1}$ and $q\geq z_{N_{0}-1}$ and also such
that $\left\vert V(p)\right\vert \leq \left\vert V(z_{1})\right\vert $ and $%
\left\vert V(q)\right\vert \leq \left\vert V(z_{N_{0}-1})\right\vert $. Let
us first consider the \textquotedblleft favorable\textquotedblright\ case
(which obtains for a sufficiently attractive potential): $z_{1}\leq r_{\text{%
min}}$ and $z_{N_{0}-1}\geq r_{\text{min}}$. In this case, we integrate (\ref%
{EqEtaZero}) from $b_{1}$ to $z_{1}$ and since in this interval both $%
V^{\prime }(r)/\left\vert V(r)\right\vert $ and $\sin \left[ 2\eta (r)\right]
$ are \textit{negative}, we infer 
\begin{equation}
\eta (z_{1})-\eta (b_{1})=\frac{\pi }{2}\leq \int_{b_{1}}^{z_{1}}dr\,\left[%
-V^{(-)}(r)\right] ^{1/2},
\end{equation}
hence \textit{a fortiori }(see (\ref{LoLimToZ1})) 
\begin{equation}
\int_{0}^{z_{1}}dr\,\left[ -V^{(-)}(r)\right] ^{1/2}>\frac{\pi }{2}.
\label{rz1}
\end{equation}
If we define $p$ via the formula 
\begin{equation}
\int_{0}^{p}dr\,\left[ -V^{(-)}(r)\right] ^{1/2}=\frac{\pi }{2},
\label{defp}
\end{equation}
then we conclude (by comparing (\ref{rz1}) with (\ref{defp})) that $p<z_{1}$%
. Moreover, since we have supposed $z_{1}\leq r_{\text{min}}$, we have also $%
\left\vert V(p)\right\vert <\left\vert V(z_{1})\right\vert $. We then
integrate (\ref{EqEtaZero}) from $z_{N_{0}-1}$ to $b_{N_{0}},$ and taking
advantage of the fact that in this interval both $V^{\prime }(r)/\left\vert
V(r)\right\vert $ and $\sin \left[ 2\,\eta (r)\right] $ are \textit{positive}%
, we infer 
\begin{equation}
\eta (b_{N_{0}})-\eta (z_{N_{0}-1})=\frac{\pi }{2}\leq
\int_{z_{N_{0}-1}}^{b_{N_{0}}}dr\,\left[ -V^{(-)}(r)\right] ^{1/2},
\end{equation}
hence \textit{a fortiori }(see (\ref{UpLimToZ(N-1)})) 
\begin{equation}
\frac{\pi }{2}<\int_{z_{N_{0}-1}}^{\infty }dr\,\left[ -V^{(-)}(r)\right]%
^{1/2}.
\end{equation}
Analogously, if we define $q$ via the formula 
\begin{equation}
\int_{q}^{\infty }dr\,\left[ -V^{(-)}(r)\right] ^{1/2}=\frac{\pi }{2},
\label{defq}
\end{equation}
we conclude that $q>z_{N_{0}-1}$. Moreover, since we have supposed $%
z_{N_{0}-1}\geq r_{\text{min}}$, we have also $\left\vert V(q)\right\vert
<\left\vert V(z_{N_{0}-1})\right\vert $. Thus if these two relations, $%
z_{1}\leq r_{\text{min}}$ and $z_{N_{0}-1}\geq r_{\text{min}},$ hold, we
obtain

\begin{equation}
N_{0}<\frac{1}{\pi }\int\nolimits_{0}^{\infty }dr\,\left[ -V^{(-)}(r)\right]%
^{1/2}+\frac{1}{4\pi }\log \left\{ \frac{\left[ V^{(-)}(r_{\text{min}})%
\right] ^{2}}{V^{(-)}(p)V^{(-)}(q)}\right\} +1,
\end{equation}
where we have used the equations (\ref{defp}) and (\ref{defq}). This last
relation imply the validity of the relation (\ref{NUL2bis}).

We need now to consider the cases where $z_{1}>r_{\text{min}}$ or $%
z_{N-1}<r_{\text{min}}$. In these cases we could have, for example, $p<z_{1}$
\textit{and} $\left\vert V(p)\right\vert >\left\vert V(z_{1})\right\vert $.

Firstly, suppose $z_{1}>r_{\text{min}}$ which implies that $z_{N_{0}-1}>r_{%
\text{min}}$. Now we impose a first condition for the applicability of the
limit NUL2: $p\leq r_{\text{min}}$. This condition is always true for a
potential which possesses enough bound states; in practice, the limit NUL2
will be applicable only when the attractive strength of the potential is
large enough. This condition ensures that $p<z_{1}$ (this is not necessarily
true, with the definition (\ref{defp}), when $z_{1}>r_{\text{min}}$).
Indeed, if $z_{1}\leq r_{\text{min}}$, we have proved it above, and if $%
z_{1}>r_{\text{min}}$, this is still true since $p\leq r_{\text{min}}$.
Moreover, since $z_{1}>r_{\text{min}}$, we have $\left\vert
V(z_{1})\right\vert >\left\vert V(z_{N_{0}-1})\right\vert >\left\vert
V(q)\right\vert \geq M$, with 
\begin{equation}
M=\min (\left\vert V(p)\right\vert ,\left\vert V(q)\right\vert )=\min \left[%
-V^{(-)}(p),-V^{(-)}(q)\right] .  \label{Min}
\end{equation}
This implies the validity of the relation (\ref{NUL2bis}).

Secondly, suppose $z_{N_{0}-1}<r_{\text{min}}$ which obviously implies $%
z_{1}<r_{\text{min}}$. Now we impose a second condition for the
applicability of the limit: $q\geq r_{\text{min}}$. This condition is always
true for a potential which possesses enough bound states. This condition
ensure that $q>z_{N_{0}-1}$ (this is not necessarily true, with the
definition (\ref{defq}), when $z_{N_{0}-1}<r_{\text{min}}$). Indeed, if $%
z_{N_{0}-1}\geq r_{\text{min}}$, we have proved it above, and if $%
z_{N_{0}-1}<r_{\text{min}}$, this is still true since $q\geq r_{\text{min}}$%
. Moreover, since $z_{N_{0}-1}<r_{\text{min}}$, we have $\left\vert
V(z_{N_{0}-1})\right\vert >\left\vert V(z_{1})\right\vert >\left\vert
V(p)\right\vert \geq M$, with $M$ defined again by (\ref{Min}). This
implies, via the definitions (\ref{S}) and (\ref{Vminus}), the validity of
the relation (\ref{NUL2bis}), and concludes our proof of the \textit{new}
upper limit NUL2.

The proof of the \textit{new }lower limit NLL2, see (\ref{NLL2bis}), is
completely analogous, except that one integrates the ODE (\ref{EqEtaZero})
from $p$ to $q$ 
\begin{equation}
\eta (q)-\eta (p)\geq \int_{p}^{q}dr\,\left[ -V^{(-)}(r)\right] ^{1/2}-\frac{%
1}{4}\int_{p}^{r_{\text{min}}}dr\,\frac{V^{\prime }(r)}{V(r)}+\frac{1}{4}%
\int_{r_{\text{min}}}^{q}dr\,\frac{V^{\prime }(r)}{V(r)},
\end{equation}%
and from the inequalities (actually valid for any positive radius, see (\ref%
{Etauplo})) 
\begin{equation}
\eta (p)\geq 0,
\end{equation}%
\begin{equation}
\eta (q)<\left( N_{0}+\frac{1}{2}\right) \,\pi ,
\end{equation}%
we clearly infer 
\begin{equation}
\eta (q)-\eta (p)<\left( N_{0}+\frac{1}{2}\right) \,\pi .
\end{equation}%
Hence 
\begin{equation}
\left( N_{0}+\frac{1}{2}\right) \,\pi >\int_{p}^{q}dr\,\left[ -V^{(-)}(r)%
\right] ^{1/2}-\frac{1}{4}\int_{p}^{r_{\text{min}}}dr\,\frac{V^{\prime }(r)}{%
V(r)}+\frac{1}{4}\int_{r_{\text{min}}}^{q}dr\,\frac{V^{\prime }(r)}{V(r)},
\end{equation}%
hence, via the definitions (\ref{defp}) and (\ref{defq}) of $p$ and $q$, 
\begin{equation}
\left( N_{0}+\frac{3}{2}\right) \,\pi >\int_{0}^{\infty }dr\,\left[
-V^{(-)}(r)\right] ^{1/2}-\frac{1}{4}\,\log \left\{ \frac{\left[ V^{(-)}(r_{%
\text{min}})\right] ^{2}}{V^{(-)}(p)\,V^{(-)}(q)}\right\} .
\label{NLL2stringent}
\end{equation}%
Note that this inequality is true in any case, provided $p\leq r_{\text{min}%
}\leq q$, and of course it implies (again, via the definitions (\ref{S}) and
(\ref{Vminus}), as well as (\ref{Min})) the validity of the marginally less
stringent lower bound (\ref{NLL2bis}) (we preferred to display in Section %
\ref{sec1} the lower bound (\ref{NLL2bis}) rather than the more stringent
one implied by (\ref{NLL2stringent}) to underline its analogy with the upper
bound (\ref{NUL2bis})).

\subsection{Proof of the lower limits NLL3s and NLL3}

\label{sec3.4}

Let us now proceed and prove (following \cite{BC}) the \textit{new }lower
limits NLL3s and NLL3, see (\ref{NLL2s}) and (\ref{NLL2}). The proof is
analogous to the proofs of the NUL2 and NLL2 limits given in the previous
subsection except that now instead of considering the equation (\ref%
{EqEtaZero}) for $\eta (r)$ we use the equation (\ref{EqEta1}). To obtain
NLL3s, we integrate the ODE (\ref{EqEta1}) from $p$ to an arbitrary radius $%
s\geq r_{\text{min}}$:

\begin{eqnarray}
\eta (s)-\eta (p) &=&\int_{p}^{s}dr\,\left[ -V^{(-)}(r)\right] ^{1/2}-\frac{1%
}{4}\int_{p}^{r_{\text{min}}}dr\,\frac{V^{\prime }(r)}{\left\vert
V(r)\right\vert }\,\sin \left[ 2\eta (r)\right]  \notag \\
&&-\frac{1}{4}\int_{r_{\text{min}}}^{s}dr\,\frac{V^{\prime }(r)}{\left\vert
V(r)\right\vert }\,\sin \left[ 2\eta (r)\right] -\int\nolimits_{p}^{s}dr\, 
\frac{\ell }{r}\,\sin \left[ 2\eta (r)\right] .
\end{eqnarray}
The right-hand side of this last equation can be minorized (since $\eta
(s)<(N_{\ell }+1/2)\pi $ and $\eta (p)>0$, see (\ref{Etauplo})) to yield

\begin{equation}
\left( N_{\ell }+\frac{1}{2}\right) \,\pi >\eta (s)-\eta (p)\geq
\int_{p}^{s}dr\,\left[ -V^{(-)}(r)\right] ^{1/2}-\frac{1}{4}\int_{p}^{r_{%
\text{min}}}dr\,\frac{V^{\prime }(r)}{V(r)}+\frac{1}{4}\int_{r_{\text{min}%
}}^{s}dr\,\frac{V^{\prime }(r)}{V(r)}-\ell \,\log \left( \frac{s}{p}\right) .
\end{equation}%

From the definition of $p$, see (\ref{P}), we finally obtain 
\begin{equation}
\left( N_{\ell }+1\right) \pi >\int_{0}^{s}dr\,\left[- V^{(-)}(r)\right]
^{1/2}-\frac{1}{4}\log \left\{ \frac{\left[ V^{(-)}(r_{\text{min}})\right]
^{2}}{V^{(-)}(p)V^{(-)}(s)}\right\} -\ell \log \left( \frac{s}{p}\right) ,
\end{equation}%
which coincides with the lower limit NLL3s, see (\ref{NLL2s}), that is
thereby proven.

To get the lower limit NLL3 we proceed as above, except that we integrate
from $p$ to $q,$ see (\ref{P}) and (\ref{Q}).

\subsection{Proof of the results in terms of comparison potentials (see
Section \protect\ref{sec1.4})}

\label{sec3.5}

Let us now prove the relations (\ref{NHlow}) and (\ref{NHup}). We assume for
this purpose that the potential $V(r)$ satisfy the negativity condition (\ref%
{V2}), but we require no monotonicity condition on $V(r)$; we do however
require the potential $V(r)$ to be nonsingular for $0\leq r<\infty $ and to
satisfy the conditions, see (\ref{VatInf}) and (\ref{VatZero}), that are
sufficient to guarantee that the quantity $S$, see (\ref{S}), be finite.

Let us now replace the potential $V(r)$ with $V(r)+H_{\lambda }^{(\ell )}(r)$%
, so that the radial Schr\"{o}dinger equation, see (\ref{ZEScha}), read now 
\begin{equation}
u^{\prime \prime }(r)=\left[ V(r)+H_{\lambda }^{(\ell )}(r)+\frac{\ell
\,(\ell +1)}{r^{2}}\right] \,u(r)
\end{equation}%
and the relation (\ref{DefVmin}) read now 
\begin{equation}
U(r)+W(r)=V(r)+H_{\lambda }^{(\ell )}(r)+\frac{\ell \,(\ell +1)}{r^{2}}.
\label{DefW}
\end{equation}%
Let us moreover set (see (\ref{DefEta}))

\begin{equation}
U(r)=V(r),  \label{setU}
\end{equation}

\begin{equation}
f(r)=\frac{V^{\prime }(r)}{4V(r)},  \label{setf}
\end{equation}%
so that (see (\ref{DefW})) 
\begin{equation}
W(r)=H_{\lambda }^{(\ell )}(r)+\frac{\ell \,(\ell +1)}{r^{2}},
\end{equation}%
with the additional requirement

\begin{equation}
H_{\lambda }^{(\ell )}(r)+\frac{\ell (\ell +1)}{r^{2}}-\left[ f(r)\right]
^{2}+f^{\prime }(r)=\beta \,\left\vert V(r)\right\vert ,  \label{fW2}
\end{equation}%
where $\beta <1$. [Note that we imposed in (\ref{fW}) that the left-hand
side of (\ref{fW2}) be positive. Actually this restriction, which was
introduced to prove that $\eta ^{\prime }(b_{n})\geq 0$ (see (\ref%
{EtaAtOddPi})), was too strong for our needs. Indeed, one can verify from (%
\ref{EqEta}) that we still have $\eta ^{\prime }(b_{n})\geq 0$ provided $%
\beta <1$]. It is easily seen that this entails for $H_{\lambda }^{(\ell
)}(r)$ the definition (\ref{DefH}) via the assignment 
\begin{equation}
\beta =1-4\lambda ^{2}.  \label{Lambda}
\end{equation}%
Hence these assignments imply that the definition of $\eta (r),$ see (\ref%
{DefEta}), reads now 
\begin{equation}
\left[ -V(r)\right] ^{1/2}\cot \left[ \eta (r)\right] =\frac{V^{\prime }(r)}{%
4\,V(r)}+\frac{u^{\prime }(r)}{u(r)},  \label{DefEta1}
\end{equation}%
and, most importantly, that the equation (\ref{EqEta}) satisfied by $\eta (r)
$ becomes now simply

\begin{equation}
\eta ^{\prime }(r)=\left\vert V(r)\right\vert ^{1/2}\left\{ 1-\beta \,\sin
^{2}\left[ \eta (r)\right] \right\} ,
\end{equation}
entailing

\begin{equation}
\eta ^{\prime }(r)\left\{ 1-\beta \,\sin ^{2}\left[ \eta (r)\right] \right\}
^{-1}=\left\vert V(r)\right\vert ^{1/2}.  \label{NewEqEta}
\end{equation}

Both sides of this last equation are now easily integrated, the right-hand
side from $r=0$ to $r=\infty ,$ and the left-hand side, correspondingly,
from $\eta =0$ (see (\ref{EtaZero}), which is clearly implied by (\ref%
{DefEta1}) and by (\ref{B0})) to $\eta (\infty )$, yielding (see (\ref{S})
and (\ref{Lambda})) 
\begin{equation}
\eta (\infty )=\,\lambda \,S\,\pi .  \label{EtaInf1}
\end{equation}
It is on the other hand clear that in this case as well 
\begin{equation}
N_{\ell }\,\pi \leq \eta (\infty )<(N_{\ell }+1)\,\pi .
\end{equation}
[Indeed, while in this case the relation (\ref{condU}) does not hold and
therefore neither (\ref{EtaInf}) nor (\ref{IneqEta}) need be true, the
relation (\ref{EtaAtEvenPi}) is still implied by the definition (\ref%
{DefEta1}), and moreover (\ref{NewEqEta}) clearly implies $\eta
^{\prime}(z_{n})\geq 0,$ entailing validity of these inequalities]. Hence
(see (\ref{EtaInf1})) 
\begin{equation}
N_{\ell }^{(V+H_{\lambda }^{(\ell )})}=\left\{ \left\{ \lambda
\,S\right\}\right\} \,,  \label{NH2}
\end{equation}
where as usual the double brace denote the integer part. In this last
formula the notation $N_{\ell }^{(V+H_{\lambda }^{(\ell )})}$ denotes of
course the number of $\ell $-wave bound states possessed by the potential $%
V(r)+H_{\lambda }^{(\ell )}(r).$

But if the \textquotedblleft additional potential\textquotedblright\ $%
H_{\lambda }^{(\ell )}(r)$, see (\ref{DefH}), is nowhere negative, this
potential $V(r)+H_{\lambda }^{(\ell )}(r)$ cannot possess less ($\ell $%
-wave) bound states than the potential $V(r)$, hence the lower limit (\ref%
{NHlow}) is proved. And under the same conditions, if the function $%
H_{\lambda }^{(\ell )}(r)$ is nowhere positive, the potential $%
V(r)+H_{\lambda }^{(\ell )}(r)$ has no less bound states than the potential $%
V(r)$, hence the upper limit (\ref{NHup}) is proven.

\subsection{Proof of the upper and lower limits NUL1 and NLL1}

\label{sec3.6}

Next, we prove the \textit{new }upper and lower limits NUL1 and NLL1, see (%
\ref{NUL1}) and (\ref{NLL1}). To prove them we of course assume the
potential $V(r)$ to possess the properties (\ref{Vr+-}), and we set $\ell =0$%
. We moreover set (see (\ref{DefEta})) 
\begin{equation}
f(r)=0
\end{equation}%
and 
\begin{equation}
U(r)=-a^{2},  \label{UeqA2}
\end{equation}%
where $a$ is a \textit{positive} constant, $a>0$, the value of which we
reserve to assign at our convenience later. Note that in this case our
assignment for $U(r)$ does not satisfy the conditions (\ref{UatInf}) and (%
\ref{UatZero}), and that the definition (\ref{DefEta}) of $\eta (r)$ now
reads 
\begin{equation}
a\,\cot \left[ \eta (r)\right] =\frac{u^{\prime }(r)}{u(r)}.
\label{DefEta1b}
\end{equation}%
Consistently with these assignments we also set (see (\ref{DefVmin})) 
\begin{equation}
W(r)=V(r)+a^{2},
\end{equation}%
and the equation satisfied by $\eta (r)$, see (\ref{EqEta}), now reads 
\begin{equation}
\eta ^{\prime }(r)=a\,\cos ^{2}\left[ \eta (r)\right] -a^{-1}\,V(r)\,\sin
^{2}\left[ \eta (r)\right] .
\end{equation}

We then integrate this ODE from $r=r_{-}$ to $r=r_{+}$: 
\begin{equation}
\eta (r_{+})-\eta (r_{-})=\int_{r_{-}}^{r_{+}}dr\,\left\{ a\,\cos ^{2}\left[%
\eta (r)\right] +a^{-1}\,\left\vert V(r)\right\vert \,\sin ^{2}\left[ \eta
(r)\right] \right\} .  \label{Eta+-}
\end{equation}
Note that we used (\ref{Vr+-b}).

Now the definition (\ref{DefEta1b}) of $\eta (r)$ entails that the radii $%
b_{n}$, see (\ref{EtaAtOddPi}), coincide with the extrema of the zero-energy
wave function $u(r),$ $u^{\prime }(b_{n})=0.$ We therefore can use (\ref%
{Vr+-a}) to conclude that, since the zero-energy wave function $u(r)$ is
concave in the interval $0\leq r<r_{-},$ the first extremum $b_{1}$ must
occur after $r_{-}$, $r_{-}<b_{1}$, hence (see (\ref{EtaAtOddPi})) 
\begin{equation}
0\leq \eta (r_{-})<\frac{\pi }{2}.
\end{equation}
Likewise, (\ref{Vr+-c}) entails that $u(r)$ is concave in the interval $%
r_{+}<r<\infty ,$ hence the last extremum, $b_{N_{0}},$ must occur before $%
r_{+}$, $b_{N_{0}}<r_{+}$, while of course $\eta (r)$ can never reach $%
(N_{0}+1)\,\pi $ (note that in this case the condition (\ref{condU}) does
not hold hence the more stringent condition (\ref{Etauplo}) does not apply).
Hence 
\begin{equation}
\left( N_{0}-\frac{1}{2}\right) \,\pi <\eta (r_{+})<\left( N_{0}+1\right)
\,\pi ,
\end{equation}
where we are of course denoting as $N_{0}$ the number of bound states
possessed by the potential $V(r)$. Hence we may assert that 
\begin{equation}
\left( N_{0}-1\right) \,\pi <\eta (r_{+})-\eta (r_{-})<\left( N_{0}+1\right)
\,\pi .  \label{etaineq}
\end{equation}

From the left-hand one of these two inequalities, and (\ref{Eta+-}), we
immediately get 
\begin{equation}
\left( N_{0}-1\right) \,\pi <\int_{r_{-}}^{r_{+}}dr\,\left\{
a\,+a^{-1}\,\left\vert V(r)\right\vert \right\} ,
\end{equation}%
since the replacement of $\cos ^{2}\left[ \eta (r)\right] $ and$\,\ \sin ^{2}%
\left[ \eta (r)\right] $ by unity in the right-hand side of (\ref{Eta+-})
entails a (further) majorization. Hence 
\begin{equation}
N_{0}<1+\frac{1}{\pi }\left[ a\,(r_{+}-r_{-})+a^{-1}\int_{0}^{\infty}dr\,\left[- V^{(-)}(r)\right] \right] ,
\end{equation}%
and by setting 
\begin{equation}
a=(r_{+}-r_{-})^{-1/2}\left( \int_{0}^{\infty}dr\,\left[-V^{(-)}(r)\right] \right) ^{1/2}
\end{equation}%
we get 
\begin{equation}
N_{0}<1+\frac{2}{\pi }\left[ \,(r_{+}-r_{-})\int_{0}^{\infty}dr\,\left[- V^{(-)}(r)\right] \right] ^{1/2}.
\end{equation}%
The result NUL1, see (\ref{NUL1}), is thereby proven.

To prove the lower limit NLL1, see (\ref{NLL1}), we use the second
inequality of (\ref{etaineq}) to get from (\ref{Eta+-}) the inequality: 
\begin{equation}
(N_{0}+1)\,\pi >\eta (r_{+})-\eta (r_{-})\geq \int_{0}^{\infty}dr\,\min 
\left[ a,a^{-1}\,\left[- V^{(-)}(r)\right] \right] ,  \label{etapm}
\end{equation}
and this coincides with the formula (\ref{NLL1}). The lower limit NLL1 is
thereby proven.

\subsection{Unified derivation of the upper limits NUL1 and NUL2}

\label{sec3.7}

In this subsection we indicate how the derivations of the two upper limits,
NUL1 and NUL2, can be unified. This entails that, in this context, the limit
NUL2 is the optimal one. In view of the previous developments our treatment
here is rather terse.

The starting point of the treatment is the ODE 
\begin{equation}
\eta ^{\prime }(r)=\left\vert U(r)\right\vert ^{1/2}\,\cos ^{2}\left[ \eta
(r)\right] -V(r)\,\left\vert U(r)\right\vert ^{-1/2}\sin ^{2}\left[ \eta (r) %
\right] +\frac{U^{\prime }(r)}{4\,U(r)}\,\sin \left[ 2\eta (r)\right] ,
\label{etasincos}
\end{equation}
that corresponds to (\ref{EqEta}) with $\ell =0,$ $f(r)=0,$ and $%
W(r)=V(r)-U(r),$ where we always assume $U(r)$ to be nonpositive, see (\ref%
{Unonneg}), but otherwise we maintain the option to assign it at our
convenience. Clearly this formula entails 
\begin{equation}
\eta ^{\prime }(r)\leq \max \left[ \left\vert
U(r)\right\vert^{1/2}\,,-V(r)\,\left\vert U(r)\right\vert ^{-1/2}\right]
+\left\vert \frac{U^{\prime }(r)}{4\,U(r)}\right\vert
\end{equation}
hence 
\begin{equation}
\eta (r_{2})-\eta (r_{1})\leq \int_{r_{1}}^{r_{2}}dr\,\left\{ \max \left[%
\left\vert U(r)\right\vert ^{1/2}\,,-V(r)\,\left\vert U(r)\right\vert ^{-1/2}%
\right] +\left\vert \frac{U^{\prime }(r)}{4\,U(r)}\right\vert \right\}.
\label{eta2-1}
\end{equation}
It is now clear that two assignments of $U(r)$ recommend themselves. One
possibility is to assume that $U(r)$ is constant, implying that the last
term in the right-hand side of this inequality vanishes: this is indeed the
choice (\ref{UeqA2}), and it leads to the neat upper limit NUL1, see the
preceding subsection \ref{sec3.6}. The other, optimal, possibility is to
equate the two arguments of the maximum functional in the right-hand side of
this inequality, (\ref{eta2-1}), namely to make the assignment (\ref{UeqV}),
and then to proceed as in subsection \ref{sec3.4}, arriving thereby to the
upper limit NUL2 (and note that a closely analogous procedure was used in 
\cite{BC}, albeit in the simpler context of a monotonically increasing
potential).

\subsection{Proof of the upper and lower limits of second kind (see Section 
\protect\ref{sec1.5})}

\label{sec3.8}

Let us now prove the results of Section \ref{sec1.5}, beginning with the
proof of the upper limit (\ref{UpLiSecKind}). To this end let us assume
first of all that $r_{J^{(\text{up, decr)}}}^{(\text{up,decr)}}>0$ (see (\ref%
{RecUpDecr}) and (\ref{Jupdecr})), and let us then introduce the piecewise
constant comparison potential $V^{(+)}(r),$ defined as follows: 
\begin{subequations}
\label{VPlus}
\begin{equation}
V^{(+)}(r)=0\quad \text{for\quad }0\leq r<r_{J^{(\text{up, decr)}}}^{(\text{%
up,decr)}},  \label{VPlusa}
\end{equation}%
\begin{equation}
V^{(+)}(r)=V(r_{j-1}^{(\text{up,decr)}})\quad \text{for\quad }r_{j-1}^{(%
\text{up,decr)}}\leq r<r_{j}^{(\text{up,decr)}}\text{\quad with\quad }j=J^{(%
\text{up,decr)}},J^{(\text{up,decr)}}-1,\ldots ,0,  \label{VPlusb}
\end{equation}%
\begin{equation}
V^{(+)}(r)=V(r_{j}^{\text{(up,incr)}})\quad \text{for\quad }r_{j-1}^{\text{%
(up,incr)}}\leq r<r_{j}^{(\text{up,incr)}}\text{\quad with\quad }j=1,\ldots
,J^{(\text{up,incr)}},  \label{VPlusc}
\end{equation}%
\begin{equation}
V^{(+)}(r)=0\quad \text{for\quad }r_{J^{(\text{up,incr)}}}\leq r.
\label{VPlusd}
\end{equation}%
It is obvious by construction (if in doubt, draw a graph!) that 
\end{subequations}
\begin{equation}
V(r)\geq V^{(+)}(r),
\end{equation}%
hence, if we indicate with $N_{0}^{(+)}$ the number of S-wave bound states
possessed by the potential $V^{(+)}(r)$ (and of course with $N_{0}$ the
number of S-wave bound states possessed by the potential $V(r)),$ clearly 
\begin{equation}
N_{0}\leq N_{0}^{(+)}.
\end{equation}

It is moreover clear from the previous treatment, see in particular (\ref%
{DefEta3}) and (\ref{EqEtaZero}) that we now write, in self-evident
notation, as follows, 
\begin{equation}
\left\vert V^{(+)}(r)\right\vert ^{1/2}\,\cot \left[ \eta ^{(+)}(r)\right] =%
\frac{u^{(+)\prime }(r)}{u^{(+)}(r)},  \label{DefEta4}
\end{equation}
\begin{equation}
\eta ^{(+)\prime }(r)=\left\vert V^{(+)}(r)\right\vert ^{1/2}-\frac{%
V^{(+)\prime }(r)}{4\left\vert V^{(+)}(r)\right\vert }\,\sin \left[
2\eta^{(+)}(r)\right] ,  \label{ODEEta+}
\end{equation}
that the number $N_{0}^{(+)}$of S-wave bound states possessed by the
potential $V^{(+)}(r)$ can be obtained in the usual manner via the solution $%
\eta ^{(+)}(r)$, for $r\geq r_{J^{(\text{up, decr)}}}^{(\text{up,decr)}}$,
of this ODE, (\ref{ODEEta+}), characterized by the boundary condition (see (%
\ref{DefEta4}) and (\ref{VPlusa}) entailing $u(r)=u^{\prime }(0)\,r$ for $%
0\leq r\leq r_{J^{(\text{up, decr})}}^{(\text{up,decr})}$) 
\begin{equation}
\left\vert V^{(+)}\left( r_{J^{(\text{up, decr)}}}^{(\text{up,decr)}%
}\right)\right\vert ^{1/2}\,\cot \left[ \eta ^{(+)}\left( r_{J^{(\text{up,
decr)}}}^{(\text{up,decr)}}\right) \right] =\left[ r_{J^{(\text{up, decr)}%
}}^{( \text{up,decr)}}\right] ^{-1},
\end{equation}
namely 
\begin{equation}
\tan \left[ \eta ^{(+)}\left( r_{J^{(\text{up, decr)}}}^{(\text{up,decr)}%
}\right) \right] =\left[ r_{J^{(\text{up, decr)}}}^{(\text{up,decr)}}\right]%
\,\left\vert V^{(+)}\left( r_{J^{(\text{up, decr)}}}^{(\text{up,decr)}
}\right) \right\vert ^{1/2}
\end{equation}
entailing 
\begin{equation}
\eta ^{(+)}\left( r_{J^{(\text{up, decr)}}}^{(\text{up,decr)}}\right) <\frac{%
\pi }{2}.  \label{BounEta+}
\end{equation}
The number $N_{0}^{(+)}$of S-wave bound states possessed by the potential $%
V^{(+)}(r)$ is then characterized by the inequality 
\begin{equation}
N_{0}^{(+)}\leq \frac{1}{\pi }\eta ^{(+)}\left( r_{J^{(\text{up, incr)}}}^{(%
\text{up,incr)}}\right) +\frac{1}{2}.  \label{NPlusUp}
\end{equation}
[Indeed, in self-evident notation, $b_{N_{0}^{(+)}}^{(+)}\leq r_{J^{(\text{%
up, incr)}}}^{(\text{up,incr)}},$ see (\ref{VPlusd}), hence $%
\eta^{(+)}\left( b_{N_{0}^{(+)}}^{(+)}\right) \leq \eta ^{(+)}\left( r_{J^{(%
\text{up, incr)}}}^{(\text{up,incr)}}\right) $ (see (\ref{ODEEta+}) and (\ref%
{VPlusc})), and $\eta ^{(+)}\left( b_{N_{0}^{(+)}}^{(+)}\right)
=\left(N_{0}^{(+)}\,-\frac{1}{2}\right) \pi ,$ see (\ref{EtaAtOddPi})].

We now introduce another solution, $\eta ^{(++)}(r),$ of the ODE (\ref%
{ODEEta+}) in the interval $r_{J^{(\text{up, decr)}}}^{(\text{up,decr)}}\leq
r\leq r_{J^{(\text{up, incr)}}}^{(\text{up,incr)}}$, characterized by the
boundary condition 
\begin{equation}
\eta ^{(++)}\left( r_{J^{(\text{up, decr)}}}^{(\text{up,decr)}}\right) =%
\frac{\pi }{2}.  \label{BounEta++}
\end{equation}
It is then plain, from a comparison of this \textquotedblleft initial
condition\textquotedblright\ (\ref{BounEta++}) with (\ref{BounEta+}), that,
throughout this interval, $\eta ^{(++)}(r)>\eta ^{(+)}(r),$ hence in
particular $\eta ^{(++)}\left( r_{J^{(\text{up, incr)}}}^{(\text{up,incr)}%
}\right) >\eta ^{(+)}\left( r_{J^{(\text{up, incr)}}}^{(\text{up,incr)}
}\right) ,$ hence from (\ref{NPlusUp}) we infer \textit{a fortiori} 
\begin{equation}
N_{0}^{(+)}<\frac{1}{\pi }\eta ^{(++)}\left( r_{J^{(\text{up, incr)}}}^{(%
\text{up,incr)}}\right) +\frac{1}{2}.  \label{UpN+}
\end{equation}

But the function $\eta ^{(++)}(r)$ can be easily evaluated in closed form,
since \textit{de facto }it satisfies the ODE 
\begin{equation}
\eta ^{(++)\prime }(r)=\left\vert V^{(+)}(r)\right\vert ^{1/2}.
\label{EqEta++}
\end{equation}%
Indeed the second term in the right-hand side of the ODE (\ref{ODEEta+}) now
vanishes: inside the intervals in which the potential $V^{(+)}(r)$ is
constant, see (\ref{VPlus}), because its derivative $V^{(+)\prime }(r)$
vanishes; and at the boundary of these intervals, where the potential $%
V^{(+)}(r)$ is discontinuous hence its derivative $V^{(+)\prime }(r)$
features a delta-function contribution, because the term $\sin \left[ 2\eta
^{(++)}(r)\right] $ vanishes: indeed, as can be immediately verified, the
ODE (\ref{EqEta++}) with the initial condition (\ref{BounEta++}) and the
piece-wise potential (\ref{VPlus}), entails that at these boundaries, say at 
$r=r_{n}^{(+)}$ with 
\begin{eqnarray}
r_{1}^{(+)} &=&r_{J^{(\text{up, decr)}}}^{(\text{up,decr)}%
},\,r_{2}^{(+)}=r_{J^{(\text{up, decr)}}-1}^{(\text{up,decr)}},\ldots
,r_{J^{(\text{up, decr)}}}^{(+)}=r_{1}^{(\text{up,decr)}},  \notag \\
r_{J^{(\text{up, decr)}}+1}^{(+)} &=&r_{\text{min}},r_{J^{(\text{up, decr)}%
}+2}^{(+)}=r_{1}^{(\text{up,incr)}},\,r_{J^{(\text{up, decr)}%
}+3}^{(+)}=r_{2}^{(\text{up,incr)}},  \notag \\
\ldots ,r_{J^{(\text{up, decr)}}+J^{(\text{up, incr)}}}^{(+)} &=&r_{J^{(%
\text{up, incr)}}-1}^{(\text{up,incr)}},\,r_{J^{(\text{up, decr)}}+J^{(\text{%
up, incr)}}+1}^{(+)}=r_{J^{(\text{up, incr)}}}^{(\text{up,incr)}},
\label{rn+}
\end{eqnarray}%
there hold the relations 
\begin{equation}
\eta ^{(++)}(r_{n}^{(+)})=n\,\frac{\pi }{2},\,n=1,2,...,J^{(\text{up, decr)}%
}+J^{(\text{up, incr)}}+1.
\end{equation}%
This last formula entails indeed $\sin \left[ 2\eta ^{(++)}(r_{n}^{(+)})%
\right] =0,$ and moreover 
\begin{equation}
\eta ^{(++)}\left( r_{J^{(\text{up, incr)}}}^{(\text{up,incr)}}\right)
=\left( J^{(\text{up, decr)}}+J^{(\text{up, incr)}}+1\right) \,\frac{\pi }{2}%
,  \label{Eta++J}
\end{equation}%
hence, via (\ref{UpN+}), 
\begin{equation}
N_{0}<\frac{1}{2}\left( J^{(\text{up,incr})}+J^{(\text{up,decr})}+2\right) ,
\label{NUp}
\end{equation}%
consistently with the \textit{new} upper limit of the second kind, see (\ref%
{UpLiSecKind}), in the case $r_{J^{(\text{up, decr)}}}^{(\text{up,decr)}}>0$.

If instead $r_{J^{(\text{up, decr)}}}^{(\text{up,decr)}}\leq 0$, the proof
is analogous, except that the formula (\ref{VPlusa}) is now irrelevant, the
\textquotedblleft initial condition\textquotedblright\ (\ref{BounEta+}) is
replaced by 
\begin{equation}
\eta ^{(+)}(0)=0,
\end{equation}%
the \textquotedblleft initial condition\textquotedblright\ (\ref{BounEta++})
is replaced by 
\begin{equation}
\eta ^{(++)}\left( 0\right) =\frac{\pi }{2}-\frac{\left\vert V\left( r_{J^{(%
\text{up,decr)}}-1}^{(\text{up,decr)}}\right) \right\vert ^{-1/2}}{r_{J^{(%
\text{up,decr)}}-1}^{(\text{up,decr)}}\,},
\end{equation}%
in the definition (\ref{rn+}) of the radii $r_{n}^{(+)}$ the lower index in
the right-hand side is always decreased by one unit (i. e., $%
r_{1}^{(+)}=r_{J^{(\text{up, decr)}}-1}^{(\text{up,decr)}}$ and so on,
entailing $r_{J^{(\text{up, decr)}}+J^{(\text{up, incr)}}}^{(+)}=r_{J^{(%
\text{up, incr)}}}^{(\text{up,incr)}}),$ hence (\ref{Eta++J}) reads 
\begin{equation}
\eta ^{(++)}\left( r_{J^{(\text{up, incr)}}}^{(\text{up,incr)}}\right)
=\left( J^{(\text{up, decr)}}+J^{(\text{up, incr)}}\right) \,\frac{\pi }{2},
\end{equation}%
and consequently (\ref{NUp}) reads 
\begin{equation}
N_{0}<\frac{1}{2}\left( J^{(\text{up,incr})}+J^{(\text{up,decr})}+1\right) ,
\end{equation}%
consistently with (\ref{UpLiSecKind}) in the case $r_{J^{(\text{up, decr)}%
}}^{(\text{up,decr)}}\leq 0.$ The proof of the \textit{new }upper limit of
the second kind (\ref{UpLiSecKind}) is thereby completed.

The proof of the \textit{new} lower limit of the second kind, see (\ref%
{LoLiSecKind}), is analogous, and we therefore only outline it here. It is
again based on the construction of a piecewise potential $V^{(-)}(r)$ that
(in this case) maximizes the potential $V(r)$ and for which the number $%
N_{0}^{(-)}$of S-wave bound states (or rather, an upper limit to this
number) can be computed easily in closed form. We manufacture this piecewise
potential $V^{(-)}(r)$ according to the following prescriptions: 
\begin{subequations}
\begin{equation}
V^{(-)}(r)=\infty \quad \text{for\quad }0\leq r<r_{0}^{\text{(lo,incr)}},
\label{VMinusa}
\end{equation}
\begin{equation}
V^{(-)}(r)=V\left( r_{j-1}^{\text{(lo,incr)}}\right) \quad \text{for\quad }
r_{j-1}^{\text{(lo,incr)}}\leq r<r_{j}^{\text{(lo,incr)}}\text{\quad
with\quad }j=1,2,\ldots ,J^{\text{(lo,incr)}}-1,  \label{VMinusb}
\end{equation}
\begin{equation}
V^{(-)}(r)=\max \left[ V\left( r_{J^{\text{(lo,incr)}}-1}^{\text{(lo,incr)}%
}\right) ,V\left( r_{J^{\text{(lo,decr)}}-1}^{\text{(lo,decr)}}\right) %
\right] \text{\quad for\quad }r_{J^{\text{(lo,incr)}}-1}^{\text{(lo,incr)}%
}\leq r<r_{J^{\text{(lo,decr)}}-1}^{\text{(lo,decr)}},  \label{VMinusc}
\end{equation}
\begin{eqnarray}
V^{(-)}(r) &=&V\left( r_{j}^{\text{(lo,decr)}}\right) \quad \text{for\quad }%
r_{j+1}^{\text{(lo,decr)}}\leq r<r_{j}^{\text{(lo,decr)}}  \notag \\
\text{with\quad }j &=&J^{\text{(lo,decr)}}-1,\ J^{\text{(lo,decr)}}-2,\ldots
,1,0,  \label{VMinusd}
\end{eqnarray}
\begin{equation}
V^{(-)}(r)=\max_{r>r_+}[V(r)] \quad \text{for\quad }r_{0}^{\text{(lo,decr)}%
}\leq r.  \label{VMinuse}
\end{equation}
It is then obvious (draw graph if in doubt!) that this potential maximizes
the original potential $V(r),$ 
\end{subequations}
\begin{equation}
V(r)\leq V^{(-)}(r),
\end{equation}
hence that the number $N_{0}^{(-)}$of its bound states provides a lower
limit to the number $N_{0}$ of bound states of the potential $V(r)$, 
\begin{equation}
N_{0}\geq N_{0}^{(-)}.
\end{equation}

But it is also clear, on the basis of the analysis given above (and leaving
to the alert reader the task to provide the details required to turn this
argument into a rigorous proof), that each of the $J^{\text{(lo,incr)}}+J^{%
\text{(lo,decr)}}$ intervals in which the piecewise potential $V^{(-)}(r)$
is negative, see (\ref{VMinusb}), (\ref{VMinusc}) and (\ref{VMinusd}), can
accommodate \textquotedblleft half a bound state\textquotedblright\ (namely
it yields an increase by $\pi /2$ of the relevant function $\eta (r)$),
except possibly for the central interval around $r_{\text{min}}$, which can
or cannot accommodate such \textquotedblleft half a bound
state\textquotedblright\ depending whether the product of the square root of
the modulus of the potential $V^{(-)}(r)$ in that interval times the length
of that interval, 
\begin{equation}
\left[ r_{J^{\text{(lo,decr)}}-1}^{\text{(lo,decr)}}-r_{J^{\text{(lo,incr)}%
}-1}^{\text{(lo,incr)}}\right] \min \left[ \left\vert V\left( r_{J^{\text{%
(lo,incr)}}-1}^{\text{(lo,incr)}}\right) \right\vert ^{1/2},\left\vert
V\left( r_{J^{\text{(lo,decr)}}-1}^{\text{(lo,decr)}}\right)
\right\vert^{1/2}\right] ,
\end{equation}
does or does not amount to no less than $\pi /2$ (and it easy to verify that
the definition of $H$ as given after (\ref{LoLiSecKind}) entails $H=0$ in
the former case, $H=1$ in the latter: see (\ref{RecLoIncr}) and (\ref%
{RecLoDecr})). This justifies the expression in the right-hand side of (\ref%
{LoLiSecKind}), except for the additional term $-1$ appearing there, which
takes care of the two extremal intervals of the potential $V^{(-)}(r)$, see (%
\ref{VMinusa}) and (\ref{VMinuse}), each of which can at most
\textquotedblleft unbound half a bound state\textquotedblright\ (i.e., cause
a decrease of the relevant $\eta (r)$ by $\pi /2$).

\section{OUTLOOK}

\label{sec4}

With modern personal computers the calculation of the number of bound states
in a given central potential $V(r)$ is an easy numerical task (especially
using (\ref{EqEta1}) with (\ref{EtaZero}) and (\ref{EtaInf})), as well as
the numerical computation of the corresponding binding energies and
eigenfunctions. It remains however of interest to obtain neat formulas which
provide directly in terms of the potential $V(r)$ upper and lower limits for
these physical quantities -- as indeed demonstrated by the continued
attention given to these problems in the recent literature \cite%
{BlSt,BC,BC2,chad03,CMS2,CMS,hund00,hund98,lapt00,lass97}. The technique
used in this paper, and in the one that preceded it \cite{BC}, go back to
the 60's (see for instance \cite{CalBook}), yet our findings demonstrate
that it can still yield remarkably neat, and cogent, \textit{new} results.
We plan to explore in the future the applicability of this approach \cite%
{CalBook} to establish upper and lower limits to the energies of bound
states, as well as to obtain upper and lower limits on the number of bound
states (or, more generally, of discrete eigenvalues) in more general
contexts, including those spectral problems on the entire line that are
relevant for the investigation of integrable nonlinear partial differential
equations, a context in which the number of discrete eigenvalues is
generally related to the number of solitons, see for instance \cite{CD}.

\acknowledgments

Most of the results reported in this paper have been obtained during the
Scientific Gathering on Integrable Systems held from November 3rd to
December 13th, 2002, at the Centro Internacional de Ciencias (CIC) in
Cuernavaca. It is a pleasure to thank professor Thomas Seligman, director of
CIC, and his collaborators and colleagues for the pleasant hospitality and
the fruitful working environment provided by CIC. One of us (FB) would like
to thank Dr. C. Semay for checking some numerical results, FNRS for
financial support (FNRS Postdoctoral Researcher position), and the Physics
Department of the University of Rome \textquotedblleft La
Sapienza\textquotedblright\ for the warm hospitality during a two-week visit
in March 2003. Likewise, the other one (FC) wishes to thank the University
of Mons-Hainaut for the pleasant hospitality during a two-day visit in April
2003.

\end{document}